\documentclass[journal]{IEEEtran}
\usepackage{placeins}
\usepackage{makecell}
\usepackage{subfig}   

\ifCLASSINFOpdf
\else

\fi
\usepackage{cite}
\usepackage{algorithm}
\usepackage{algpseudocode}
\usepackage{amsmath,amssymb}
\usepackage{stfloats}

\usepackage{color}
\usepackage{placeins}
\usepackage{float}
\usepackage[colorlinks=true,linkcolor=black,citecolor=black,urlcolor=black]{hyperref}
\usepackage{tabularx,colortbl,multirow}
\usepackage{enumerate}
\usepackage{enumitem}
\usepackage{multirow}
\usepackage{makecell}

\usepackage{bm}
\newcommand\figref[1]{}

\renewcommand\eqref[1]{(\ref{#1})}

\usepackage{tikz}
	\usetikzlibrary{shapes,decorations,decorations.pathreplacing}
	\usetikzlibrary{arrows,shapes,positioning,calc,fadings,patterns}
	\usetikzlibrary{decorations,decorations.pathreplacing,decorations.pathmorphing}
\usepackage{tikz-3dplot}

\usepackage{booktabs} % used for the table
\usepackage{threeparttable}
\usepackage{amssymb}

\begin{document}
\title{A Hybrid Analytical-Neural Residual Surrogate for Fast Radio Propagation Prediction over Irregular Terrain}

\author{Yongyuan Hu, Jiahao Li, Keqi Ni, Yuqi Ouyang~\IEEEmembership{Member,~IEEE}, and Hao Qin~\IEEEmembership{Member,~IEEE}% <-this % stops a space
	%
% \thanks{Manuscript received XX XX, XXXX; revised XX XX, XXXX; accepted XX XX, XXXX. \textit{(Corresponding authors: Ximing Xie; Hao Qin.)}}
\thanks{Yongyuan Hu is with the Swanson School of Engineering, University of Pittsburgh, Pittsburgh, PA 15261, USA (derrick.hu@pitt.edu).}% <-this % stops a space
\thanks{Jiahao Li, Keqi Ni, and Hao Qin are with the School of Electronics and Information Engineering, Sichuan University, Chengdu 610065, China (2023141520010@stu.scu.edu.cn, 2023141520023@stu.scu.edu.cn, hao.qin@scu.edu.cn).}% <-this % stops a space
\thanks{Yuqi Ouyang is with the College of Computer Science, Sichuan University, Chengdu 610065, China (yuqi.ouyang@scu.edu.cn).}% <-this % stops a space
}

% journal name
%\markboth{IEEE ANTENNAS AND WIRELESS PROPAGATION LETTERS}%
\markboth{IEEE TRANSACTIONS ON ANTENNAS AND PROPAGATION, VOL.~XX, NO.~XX, MONTH~2026}%
{Author \MakeLowercase{\textit{et al.}}: Hybrid Analytical-Neural Residual Surrogate for Terrain Propagation}

% use for special paper notices
%\IEEEspecialpapernotice{(Invited Paper)}

\maketitle

\begin{abstract}
	Fast prediction of radio-wave propagation over irregular terrain is important for coverage analysis in mountainous environments, where many terrain and antenna configurations must be evaluated repeatedly.
	High-fidelity split-step parabolic equation (SSPE) solvers can capture terrain-induced shadowing and interference, but their computational cost is high for repeated planning studies. Learning-based surrogates reduce this cost, but direct field regression asks the network to learn both analytically tractable propagation trends and terrain-dependent residual structure.
	This paper proposes a hybrid analytical-neural residual surrogate that separates these two parts. A deterministic analytical baseline represents free-space spreading, antenna directivity, modified knife-edge diffraction, and terrain-aware specular reflection, while a compact coarse-to-fine neural network predicts the residual field. The network uses baseline-derived guidance channels, anisotropic convolutions, and feature-wise antenna conditioning to match the directional structure of terrain propagation fields.
	On synthetic SSPE data at $1.0$~GHz, the proposed model achieves an MAE of $2.1061$~dB, compared with $3.3955$~dB for a standard U-Net and $3.2560$~dB for a reproduced physics-guided CNN, while reducing the average inference time to $11.30$~ms per sample. A zero-shot test at $1.8$~GHz gives an MAE of $2.8692$~dB without retraining, and a case study on the measured Hjorringvej terrain profile shows that the prediction remains consistent with the SSPE reference and follows the measured path-loss trend.

\end{abstract}

\section{Introduction}

Accurate and efficient prediction of radio-wave propagation over
irregular terrain is a recurring requirement in wireless coverage
analysis and spectrum planning for mountainous regions. Among
high-fidelity propagation solvers, the two-way split-step parabolic
equation (SSPE) method offers a useful compromise between physical
fidelity and computational cost~\cite{guo2017pade,ozgun2009recursive,
zhang2016applying,qin2024comparative}. Its cost nevertheless remains
high for planning workflows in which terrain profiles, antenna heights,
downtilts, beamwidths, or operating frequencies must be varied many
times.

Machine learning (ML) surrogates have therefore been investigated as a
way to accelerate electromagnetic and radio-propagation
simulations~\cite{seretis2022toward,seretis2023generalizable,
bakirtzis2022deepray,9210205,9311729,10057211,10065920}. In indoor,
transportation, and other site-specific scenarios, physics-guided
convolutional models have recovered dense propagation fields with much
lower inference cost than numerical simulation~\cite{seretis2022toward,
bakirtzis2022deepray,huang2024tunnel}. For irregular terrain, Brennan
and McGuinness~\cite{brennan2023site} considered path-loss prediction,
and Huang~\textit{et al.}~\cite{huang2025generalizable} extended the
problem to two-dimensional field-map prediction using parameter
embedding, positional encoding, and a nested U-shaped backbone.

Most existing learning-based terrain surrogates still use direct
full-field regression~\cite{brennan2023site,huang2025generalizable}, a
formulation also common in related propagation models~\cite{
zhang2020cellular,liu2021deeplora,levie2021radiounet}. Under this
formulation, the network must reproduce the entire field from terrain
and antenna inputs. This places analytically predictable effects, such
as free-space spreading, antenna directivity, diffraction, and
reflection, in the same learned mapping as terrain-coupled residual
effects. For deployment settings in which a surrogate may run on
resource-constrained compute nodes~\cite{chen2021distributed}, this is
an inefficient use of model capacity. It also weakens the physical
separation between large-scale propagation trends and the residual
structure that is harder to approximate analytically, a distinction
that has proved useful in residual-learning formulations for other
inverse and restoration problems~\cite{kim2016accurate,zhang2017beyond}.

Direct regression also tends to absorb carrier-frequency dependence
implicitly into the learned weights~\cite{huang2025generalizable,
alrabeiah2019deep,sung2025commUNext}. When the operating frequency
changes, a direct surrogate usually has to be retrained or at least
fine-tuned on newly generated data. This is inconvenient in coverage
analysis, where nearby operating bands may need to be screened during
the same design process. A formulation that keeps the dominant
frequency dependence in an analytical component can reduce this
burden, provided that the learned residual remains informative outside
the training frequency.

A further issue is the mismatch between generic image backbones and
terrain-propagation fields. SSPE fields over irregular terrain have a
directional, multiscale structure tied to the dominant range direction
and the terrain boundary~\cite{levy2000parabolic,apaydin2017radio}.
Variations along range and height are not interchangeable, and antenna
height, downtilt, and beamwidth act as global controls rather than
ordinary local image channels. Standard convolutional architectures,
which were developed for visually balanced image patterns, do not
explicitly encode these features~\cite{boscaini2016learning,
yu2017dilated}. Feature-wise modulation provides a natural way to
inject global conditioning variables into intermediate feature
maps~\cite{perez2018film}.

This paper reformulates terrain-propagation prediction as a hybrid
analytical-neural residual-learning problem. A deterministic analytical
baseline is first constructed from free-space spreading, antenna
directivity, modified knife-edge diffraction, and terrain-aware
specular reflection. A neural network then predicts only the residual
correction relative to this baseline, rather than regressing the whole
SSPE field directly. In this formulation, the analytical component
captures the dominant propagation trend, while the neural component
focuses on the terrain-dependent mismatch that remains difficult to
describe with a compact closed-form model.
This differs from physics-guided direct-regression models in which
physical quantities are mainly used as auxiliary inputs; here, the
analytical field estimate is the prediction reference, and the learned
model is trained to correct it.

The proposed architecture is built around this analytical-neural
decomposition. A coarse-to-fine residual network uses anisotropic
convolutions to follow the dominant propagation direction,
baseline-derived guidance channels to expose local physical context,
and feature-wise modulation to inject antenna parameters as global
conditioning variables. Because the carrier frequency enters the
analytical component explicitly, the same learned residual model can be
evaluated at an unseen frequency by recomputing the baseline. We test
this property for transfer from $1.0$ to $1.8$~GHz without retraining.

The contributions of this work are therefore threefold. First, it
introduces a hybrid analytical-neural residual formulation for
irregular-terrain field prediction, in which an analytical baseline
removes dominant propagation structure before neural correction.
Second, it develops a lightweight coarse-to-fine architecture tailored
to the anisotropic structure of terrain-propagation fields. Third, it
evaluates the resulting surrogate against synthetic SSPE data,
reproduced learning-based baselines, an unseen-frequency test case, and
a measured Hjorringvej terrain profile.

The remainder of this paper is organized as follows. Section~II
presents the problem formulation, the deterministic baseline, and
the proposed coarse-to-fine model. Section~III reports numerical
results, including ablation studies, comparison with the reproduced
baseline method, and zero-shot frequency-transfer experiments.
Section~IV provides validation on the measured Hjorringvej terrain.
Section~V concludes the paper and discusses future work.

%%%%%%%%%%%%%%%%%%%%%%%%%%%%%%%%%%%%%%%%%%%%%%%%%%%%%%%%%%%%%%%%%%%%%%%%%%%%%%%%
%%%%%%%%%%%%%%%%%%%%%%%%%%%%%%%%%%%%%%%%%%%%%%%%%%%%%%%%%%%%%%%%%%%%%%%%%%%%%%%%
\section{ML-Based Terrain Propagation Model}
\begin{figure*}[!t]
    \centering
    \includegraphics[width=0.88\textwidth]{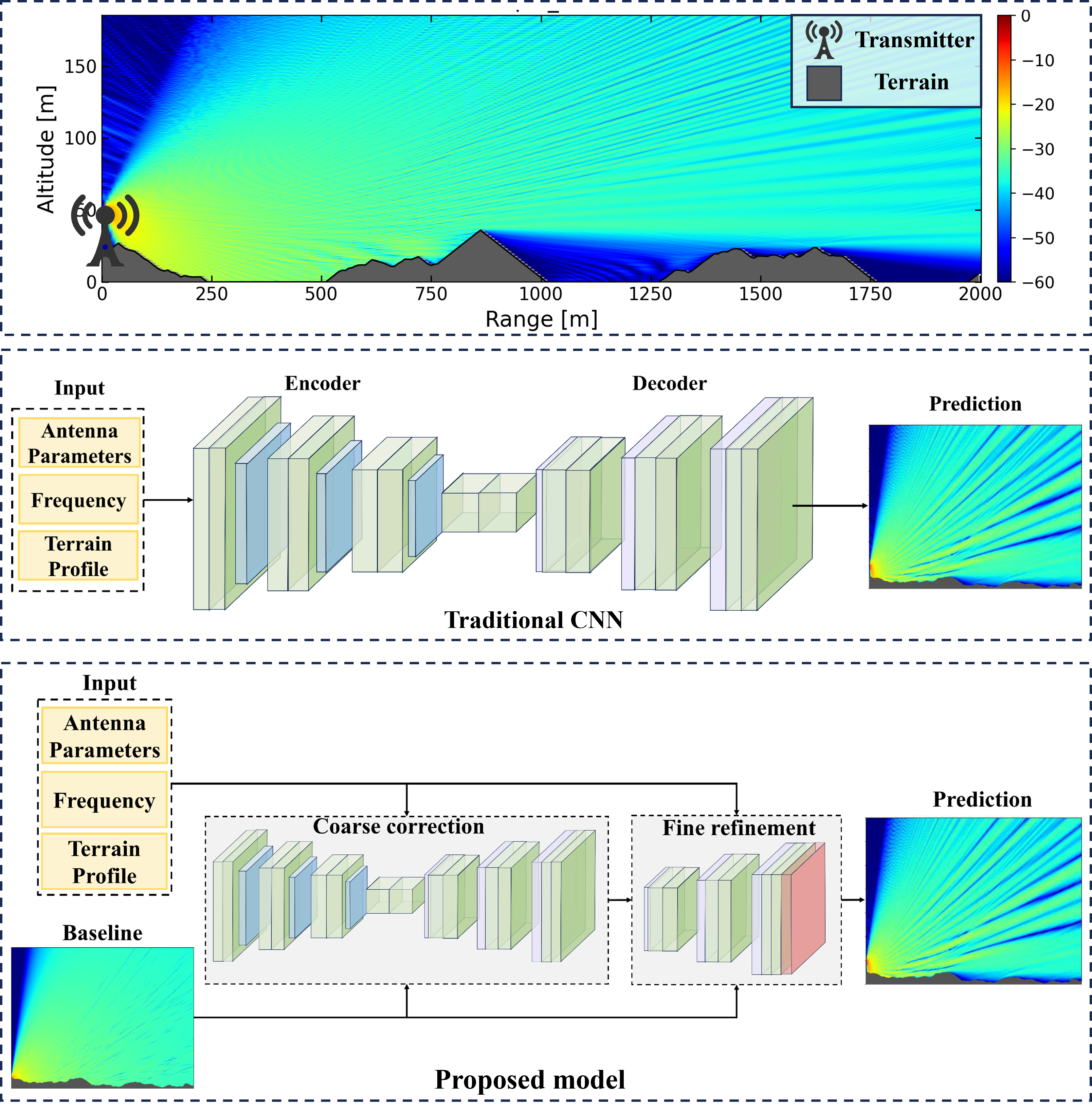}
    \caption{Overview of the propagation scenario and the proposed
hybrid analytical-neural baseline-residual framework, contrasted with
conventional full-field regression.}
    \label{fig:overview}
\end{figure*}

%加的总领性文字
This section describes the proposed hybrid analytical-neural
terrain-propagation surrogate. The model is organized around a
baseline-residual decomposition: a deterministic analytical baseline
accounts for the dominant propagation trend, and a compact neural
network predicts the remaining field correction. Fig.~\ref{fig:overview}
gives an overview of the setting and the proposed formulation. The top
panel shows an antenna radiating over an irregular terrain profile,
where the target is the two-dimensional field map defined by the terrain
geometry and antenna configuration. The middle panel illustrates the
conventional full-field regression setting. The bottom panel shows the
proposed alternative, in which the neural model learns a residual
relative to the analytical baseline. Section~II-A summarizes the SSPE
model used to generate the reference fields. Section~II-B defines the
residual formulation and analytical baseline. Section~II-C describes
the baseline-derived inputs. Section~II-D presents the coarse-to-fine
neural architecture, and Section~II-E gives the training objective.

%Figure 1: overall framework overview

%Figure 2
\begin{figure}[t]
    \centering
    \subfloat[]{%
        \includegraphics[width=0.47\columnwidth]{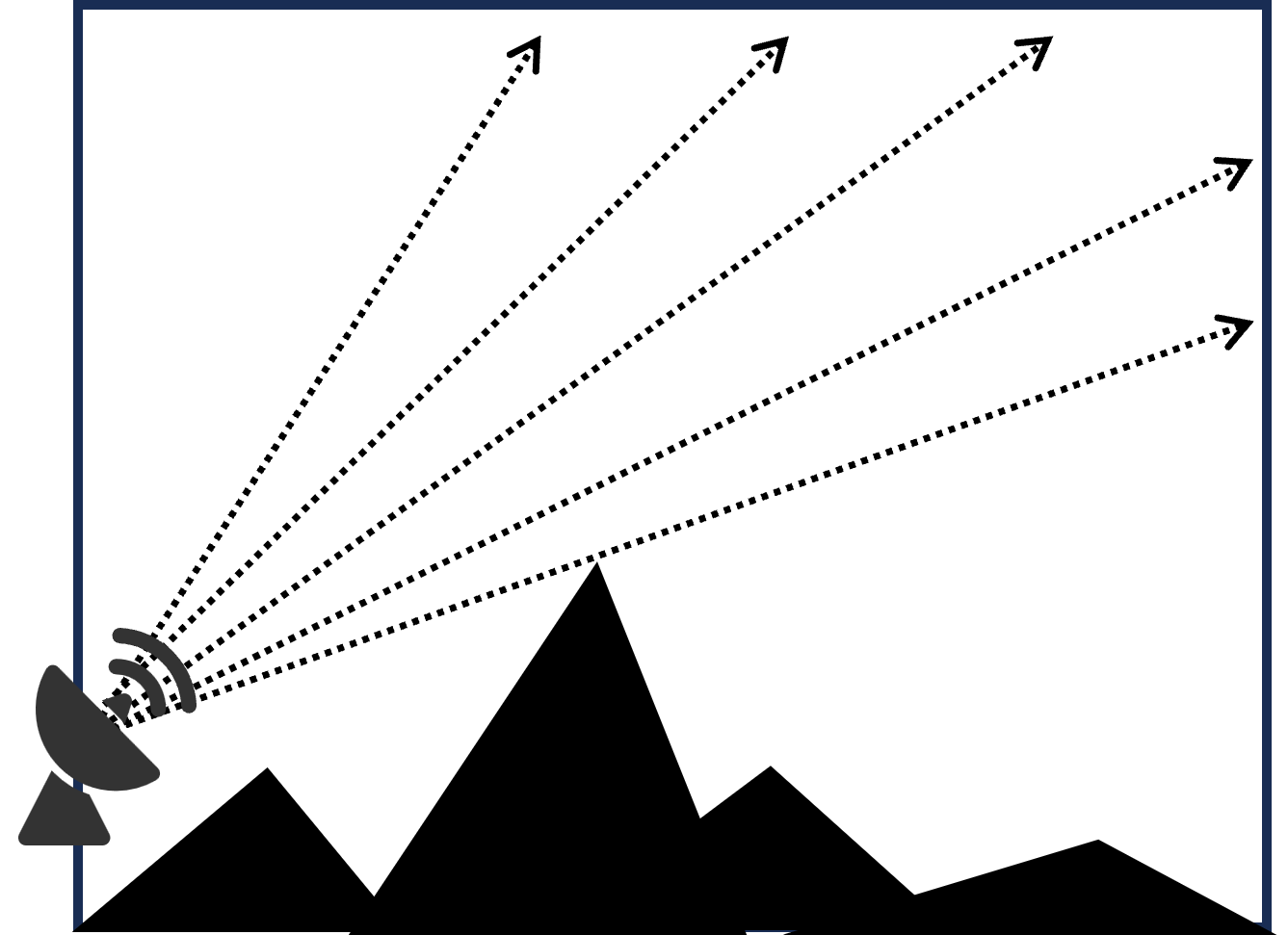}%
        \label{fig:2a}%
    }\hfill
    \subfloat[]{%
        \includegraphics[width=0.47\columnwidth]{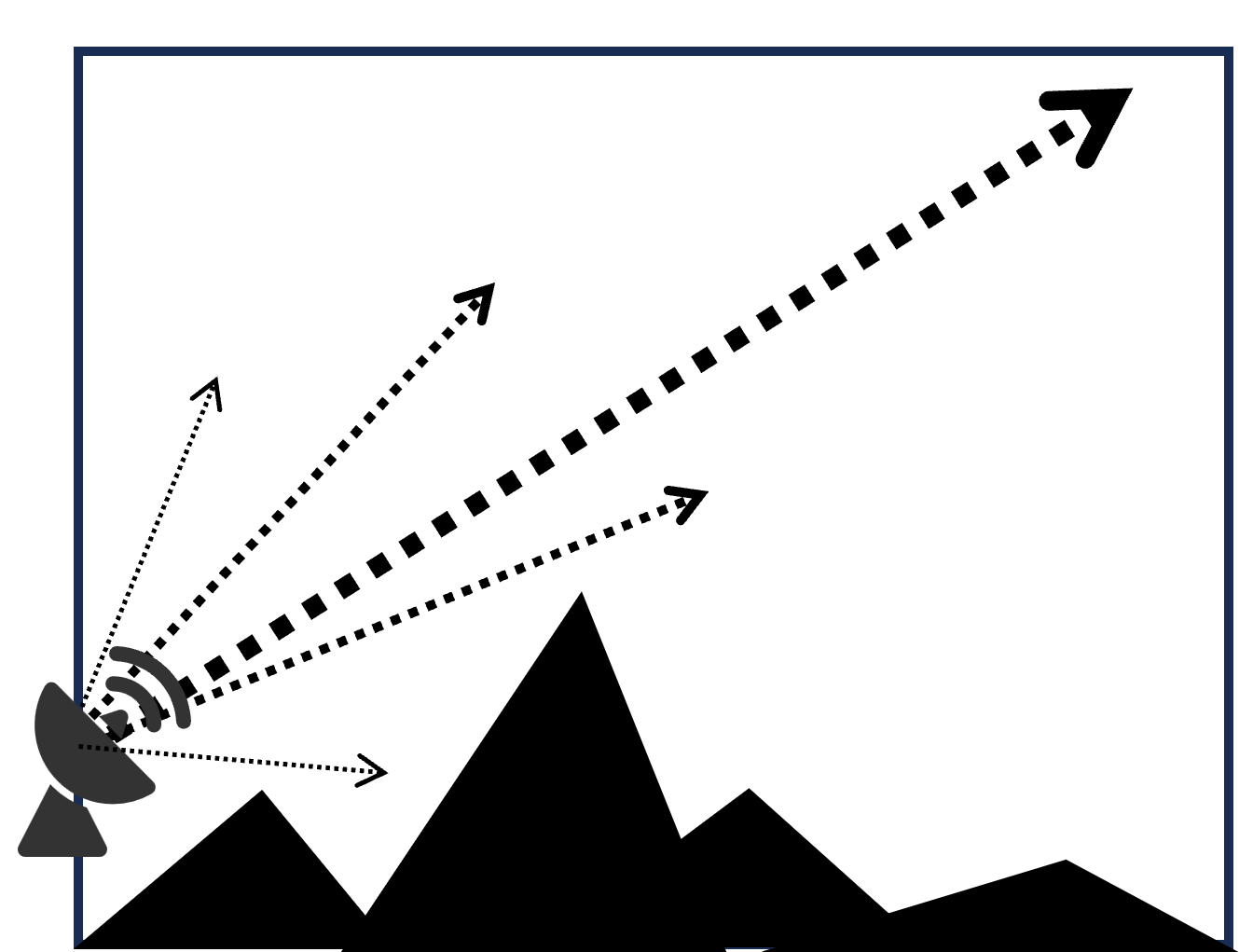}%
        \label{fig:2b}%
    }\\[0.5em]

    \subfloat[]{%
        \includegraphics[width=0.47\columnwidth]{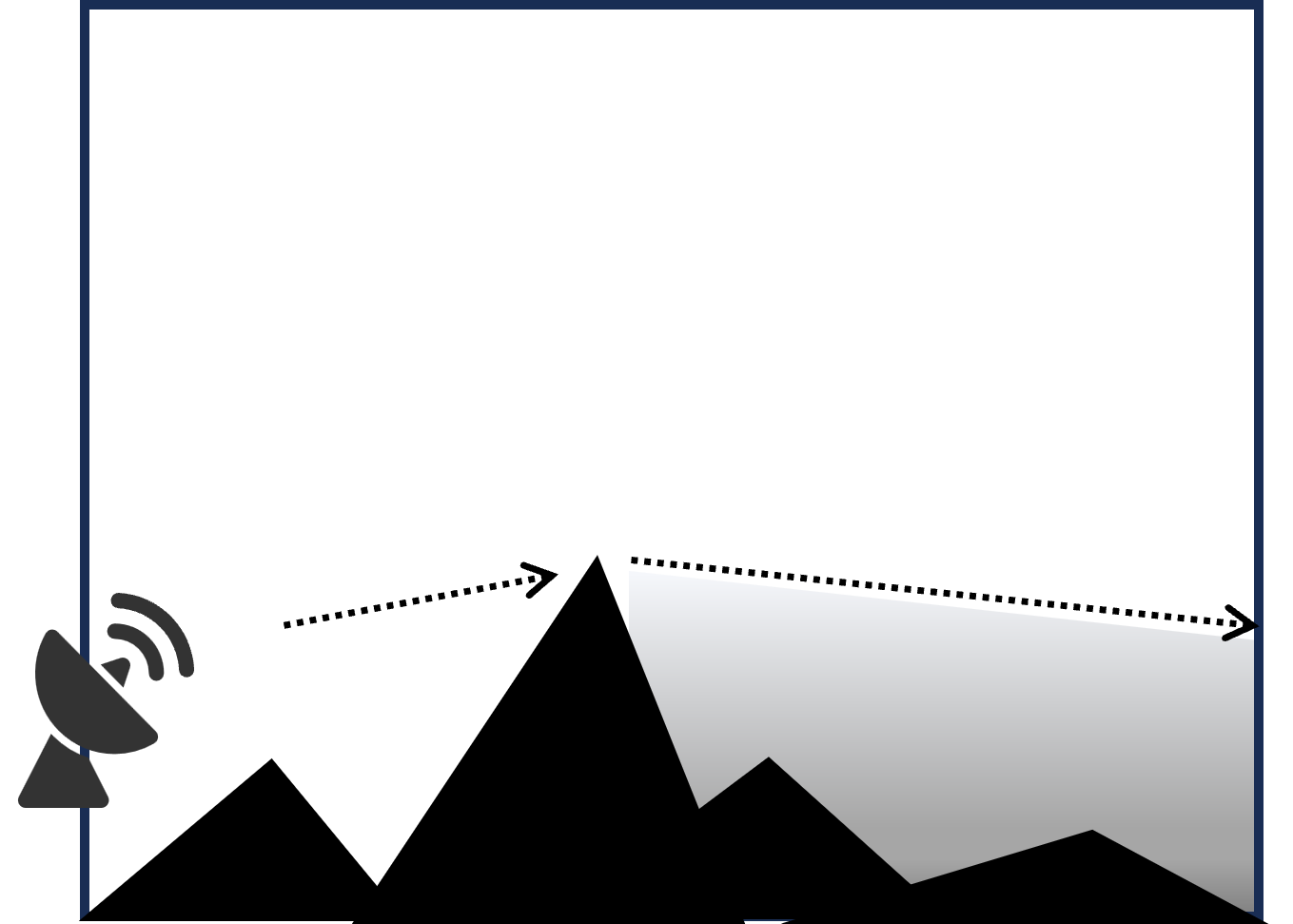}%
        \label{fig:2c}%
    }\hfill
    \subfloat[]{%
        \includegraphics[width=0.47\columnwidth]{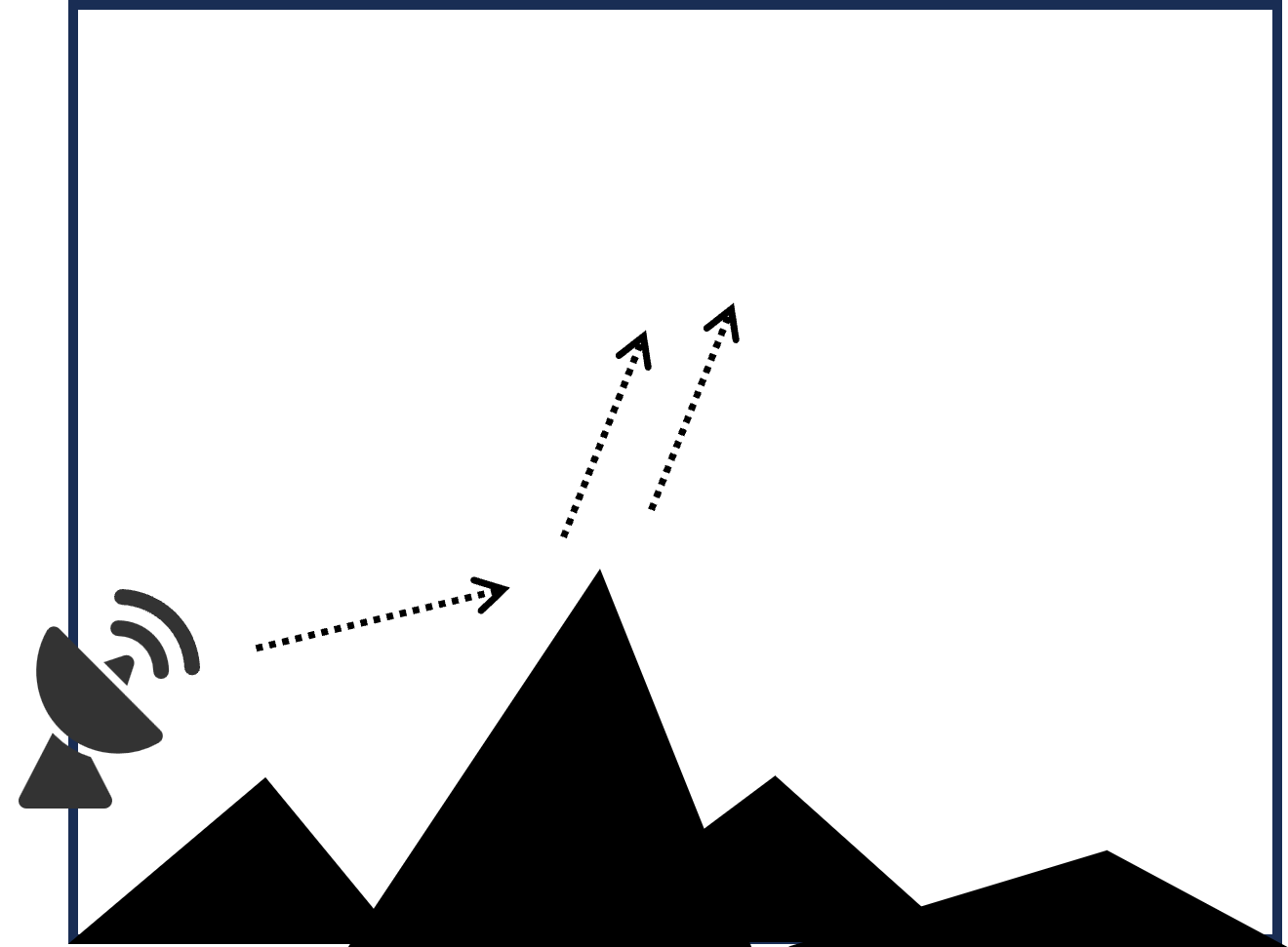}%
        \label{fig:2d}%
    }

   \caption{Schematic illustration of the four mechanisms considered in
the deterministic analytical baseline. (a) Free-space spreading.
(b) Gaussian antenna directivity. (c) Modified knife-edge diffraction.
(d) Terrain-aware specular reflection.}
    \label{fig:four_images}
\end{figure}

\subsection{Two-Way SSPE Model}

The reference data used in this work are generated with the two-way
SSPE method, a parabolic-equation approach derived from the Helmholtz
wave equation~\cite{ozgun2009recursive,apaydin2017radio,
levy2000parabolic}. The two-way formulation accounts for forward and
backward propagation components in the presence of terrain obstacles.
For a forward marching step, the reduced field in the next range plane
is obtained from
\begin{equation}
u(x, z + \Delta z) = \mathcal{F}^{-1}\!\left\{C(k_x)\,\mathcal{F}\{u(x,z)\}\right\},
\label{eq:sspe_forward}
\end{equation}
where $\mathcal{F}$ and $\mathcal{F}^{-1}$ denote the Fourier transform
pair, $k_x$ is the spectral variable, $u$ is the reduced field, and
\begin{equation}
C(k_x) = \exp\!\left(\frac{-jk_x^2\,\Delta z}{2k_0}\right),
\label{eq:sspe_propagator}
\end{equation}
where $k_0 = 2\pi f/c$ is the free-space wavenumber at carrier
frequency $f$. The backward marching step is written as
\begin{equation}
u(x, z - \Delta z) = \mathcal{F}^{-1}\!\left\{C(k_x)\,\mathcal{F}\{u(x,z)\}\right\}.
\label{eq:sspe_backward}
\end{equation}
Each SSPE simulation produces a two-dimensional complex field map over
the terrain domain. The field magnitude is converted to decibel scale
and used as the supervised learning target. The surrogate is then
trained to approximate this SSPE mapping at much lower inference cost.

\subsection{Problem Formulation and Analytical Baseline}

Let $Y\in\mathbb{R}^{H\times W}$ denote the SSPE field map in decibel
scale. Instead of predicting $Y$ directly from the terrain profile and
antenna parameters, we decompose the target as
\begin{equation}
Y(x,z)=B(x,z)+R(x,z),
\end{equation}
where $B$ is a deterministic analytical baseline and $R$ is the residual
correction learned by the network. The purpose of this decomposition is
not to make the analytical baseline fully accurate on its own, but to
remove the dominant low-complexity field trend before learning.
Thus, the baseline should be interpreted as a compact reference field
for residual learning rather than as a standalone substitute for the
SSPE solver.

The baseline is constructed in complex-field form as
\begin{equation}
E_{\mathrm{base}}(x,z)=E_{\mathrm{fspl}}(x,z)\,G_{\mathrm{ant}}(x,z)\,A_{\mathrm{ke}}(x,z)+E_{\mathrm{refl}}(x,z),
\end{equation}
where $E_{\mathrm{fspl}}$ models free-space spreading,
$G_{\mathrm{ant}}$ is the antenna directivity factor, $A_{\mathrm{ke}}$
is a modified knife-edge diffraction term, and $E_{\mathrm{refl}}$ is a
terrain-aware specular-reflection component. These four terms are shown
schematically in Fig.~\ref{fig:four_images}.

The direct free-space term is written as
\begin{equation}
E_{\mathrm{fspl}}(x,z)=d^{-1/2}e^{-jk_0 d},
\qquad
 d=\sqrt{x^2+(z-z_{\mathrm{tx}})^2},
\end{equation}
where $z_{\mathrm{tx}}$ is the transmitter height and $k_0=2\pi f/c$ is
the free-space wavenumber. The antenna pattern is approximated by a
Gaussian elevation beam,
\begin{equation}
G_{\mathrm{ant}}(x,z)=\exp\!\left[-\frac{(\theta_{\mathrm{dir}}+\theta_{\mathrm{tilt}})^2}{4\sigma_\theta^2}\right],
\qquad
\sigma_\theta=\frac{\theta_{3\mathrm{dB}}}{2\sqrt{2\ln 2}},
\end{equation}
where $\theta_{\mathrm{dir}}=\arctan((z-z_{\mathrm{tx}})/x)$, $\theta_{\mathrm{tilt}}$ is the antenna downtilt, and $\theta_{3\mathrm{dB}}$ is the 3-dB beamwidth.

Diffraction is represented by a modified single knife-edge
approximation. For each observation point, the dominant terrain
obstacle is selected as the point that maximizes the Fresnel
diffraction parameter along the transmitter--receiver path. The
reference loss follows Recommendation ITU-R P.526 and is then adjusted
to avoid unrealistic attenuation outside the shadowed region: 1) the
loss is softened by a fixed scale factor $\alpha=0.7$; 2) the
diffraction penalty is removed above an effective cap height so that
non-shadowed points are not over-attenuated; and 3) the post-peak
shadow boundary is constrained to descend monotonically with a maximum
slope of $s_{\max}=-0.01$ m/m. The resulting attenuation factor is
\begin{equation}
A_{\mathrm{ke}}(x,z)=10^{-L_{\mathrm{ke}}(x,z)/20}.
\end{equation}

The reflected component is computed with an image construction relative
to the local terrain tangent at the specular point. Its amplitude is
weighted by the Fresnel reflection coefficient for a lossy dielectric
ground. This term retains only the dominant specular contribution and
does not attempt to replace the SSPE solution.

The deterministic baseline in decibel scale is finally obtained as
\begin{equation}
B(x,z)=\operatorname{clip}_{[-200,0]}\left(20\log_{10}|E_{\mathrm{base}}(x,z)|\right).
\end{equation}
Consistent with the dataset representation, below-terrain pixels are
assigned the floor value. Because the carrier frequency enters $B$
through the analytical terms, testing the trained surrogate at another
frequency only requires recomputing the baseline; the learned weights
are kept fixed in the zero-shot experiment reported in
Section~\ref{subsec:generalization}.

%Figure 3: detailed two-stage architecture
\begin{figure*}[!t]
    \centering
    \includegraphics[width=0.88\textwidth]{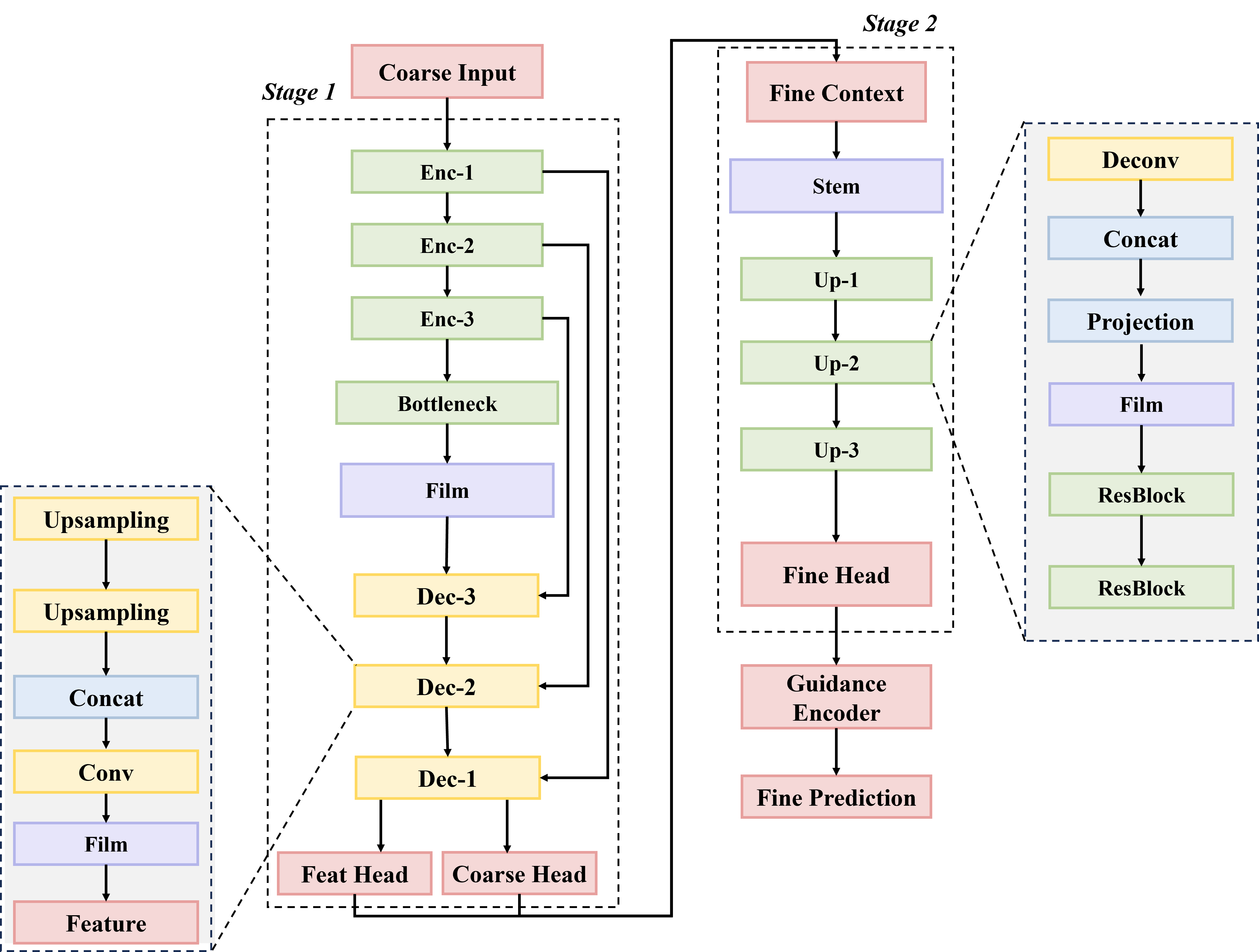}
    \caption{Detailed structure of the proposed two-stage coarse-to-fine network.}
    \label{fig:architecture}
\end{figure*}

\subsection{Baseline-Derived Input Construction}

The network input is built from the deterministic baseline and terrain
geometry, rather than from raw terrain samples alone. We first construct
a signed boundary-distance map
\begin{equation}
D_b(x,z)=\frac{z-h(x)}{\max_{x,z}|z-h(x)|},
\end{equation}
whose sign indicates whether a point lies above or below the terrain
boundary. Stage~1 operates on a coarse representation obtained by
downsampling only the height dimension by a factor of~8,
\begin{equation}
X_c=[\tilde{B}^{c},D_b^{c}].
\end{equation}
This two-channel input gives the coarse network the dominant field
trend and the terrain boundary in a compact form.

For full-resolution refinement, we additionally construct a seven-channel guidance tensor
\begin{equation}
G=[\tilde{B},M,G_z,G_x,C,X,Z],
\end{equation}
where $M$ is the validity mask, $G_z$ and $G_x$ are the normalized
vertical and horizontal gradients of the baseline, $C$ is the
normalized clearance above the terrain surface, and $X$ and $Z$ are the
normalized range and height coordinates. Antenna height, downtilt, and
beamwidth are represented by a low-dimensional conditioning vector and
injected through feature-wise linear modulation (FiLM)~\cite{
perez2018film}. These channels expose the baseline trend, terrain
boundary, local baseline gradients, and spatial position before the
nonlinear residual mapping is learned.

\subsection{Two-Stage Coarse-to-Fine Network}

The residual field is predicted in two stages. Stage~1 is a compact
U-Net-like predictor operating on the $H/8\times W$ grid. It uses
anisotropic convolutions aligned with the range axis and a dilated
bottleneck to capture long-range propagation structure at reduced
height resolution. This stage outputs a coarse residual $\hat{r}_c$
together with a compact latent feature representation.

Stage~2 reconstructs the full-resolution residual using the coarse prediction, the latent coarse features, and multiscale guidance extracted from $G$. Its decoder upsamples only along the height dimension,
\begin{equation}
H/8 \rightarrow H/4 \rightarrow H/2 \rightarrow H,
\end{equation}
which follows the asymmetric sampling of the SSPE grid and concentrates
most computation at reduced resolution. Antenna conditioning is applied
in both stages through FiLM because antenna parameters affect the field
distribution globally rather than as local spatial textures. The
overall architecture is illustrated in Fig.~\ref{fig:architecture}.

Let $\hat{r}_f$ denote the refined full-resolution residual. The final field prediction is obtained by
\begin{equation}
\hat{Y}(x,z)=B(x,z)+\sigma_f\hat{r}_f(x,z)+\mu_f,
\end{equation}
where $\mu_f$ and $\sigma_f$ are the training-set statistics of the fine
residual. The coarse stage therefore handles the dominant residual
envelope, while the full-resolution stage refines local shadow
transitions and interference details.

\subsection{Training Objective and Optimization}

Let $r_c$ and $r_f$ denote the normalized coarse and fine residual
targets, and let $\hat{r}_c$ and $\hat{r}_f$ be the corresponding
network outputs. Training uses a weighted masked Huber loss evaluated
only on valid pixels above the terrain:
\begin{equation}
\mathcal{L}=\lambda_c\,\mathcal{L}_{\mathrm{Huber}}(\hat{r}_c,r_c;M_c)
+\lambda_f\,\mathcal{L}_{\mathrm{Huber}}(\hat{r}_f,r_f;M_f),
\end{equation}
where $\lambda_c=1$ and $\lambda_f=2$. The coarse and fine residuals
are normalized separately over valid pixels only.

The network is optimized with AdamW using an initial learning rate of
$5\times10^{-4}$ and a weight decay of $10^{-4}$. A cosine-annealing
schedule is used during training, together with mixed-precision
training, gradient clipping, and early stopping.

%%%%%%%%%%%%%%%%%%%%%%%%%%%%%%%%%%%%%%%%%%%%%%%%%%%%%%%%%%%%%%%%%%%%%%%%%%%%%%%%
%%%%%%%%%%%%%%%%%%%%%%%%%%%%%%%%%%%%%%%%%%%%%%%%%%%%%%%%%%%%%%%%%%%%%%%%%%%%%%%%

%%%%%%%%%%%%%%%%%%%%%%%%%%%%%%%%%%%%%%%%%%%%%%%%%%%%%%%%%%%%%%%%%%%%%%%%%%%%%%%%
%%%%%%%%%%%%%%%%%%%%%%%%%%%%%%%%%%%%%%%%%%%%%%%%%%%%%%%%%%%%%%%%%%%%%%%%%%%%%%%%

\section{Numerical Results}\label{Section III}

% Fig4. Two-column layout to prevent float overflow: 7 sub-figures arranged as 4+3 across two columns
\begin{figure*}[!t]
    \centering
    \begin{minipage}[t]{0.49\textwidth}
        \centering
        \subfloat[]{\includegraphics[width=0.95\linewidth]{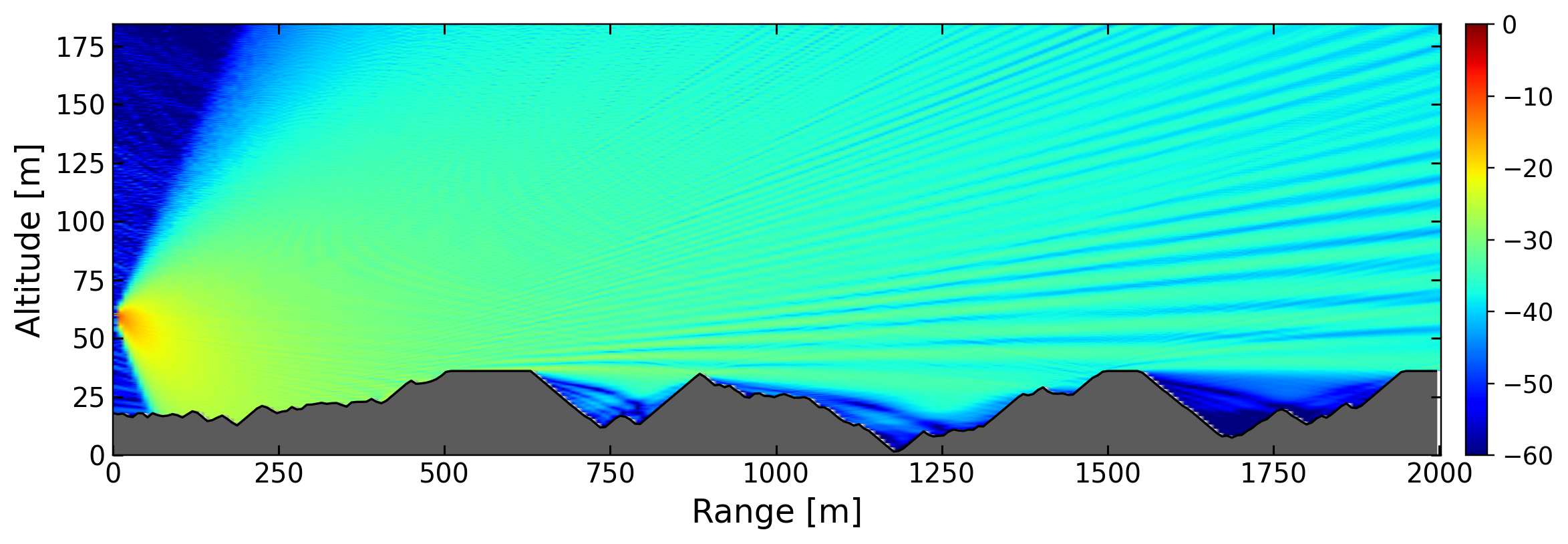}}\\[0.3ex]
        \subfloat[]{\includegraphics[width=0.95\linewidth]{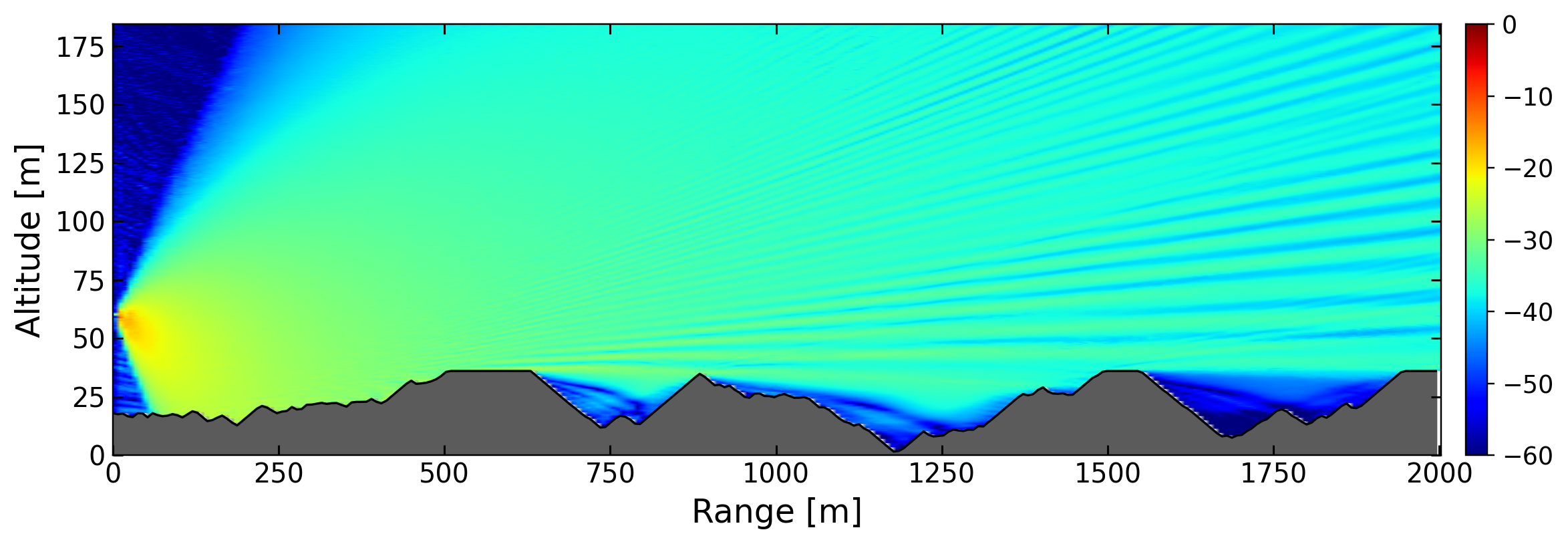}}\\[0.3ex]
        \subfloat[]{\includegraphics[width=0.95\linewidth]{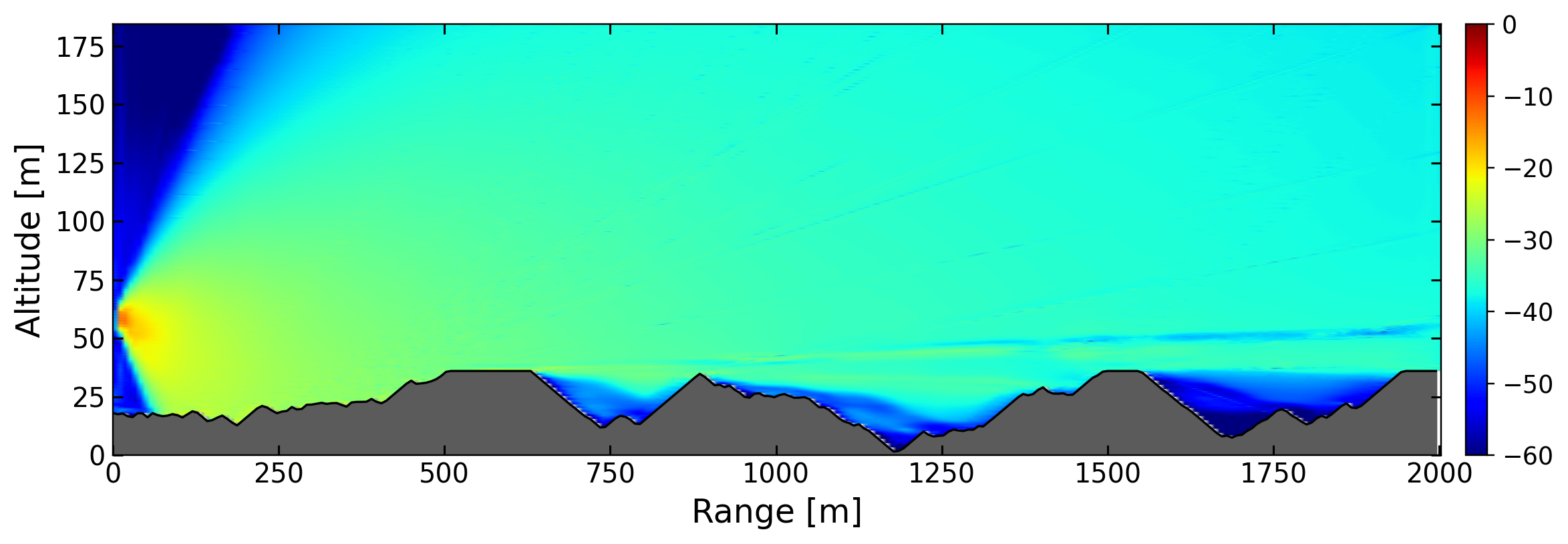}}\\[0.3ex]
        \subfloat[]{\includegraphics[width=0.95\linewidth]{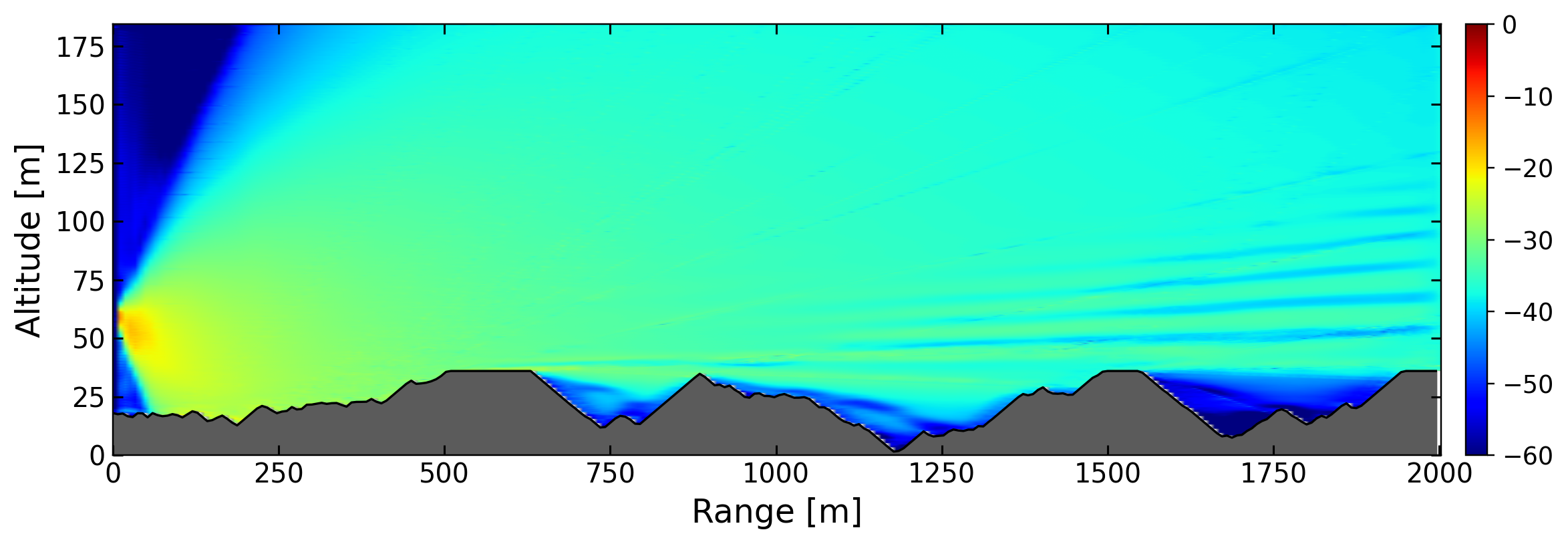}}
    \end{minipage}%
    \hfill
    \begin{minipage}[t]{0.49\textwidth}
        \centering
        \subfloat[]{\includegraphics[width=0.95\linewidth]{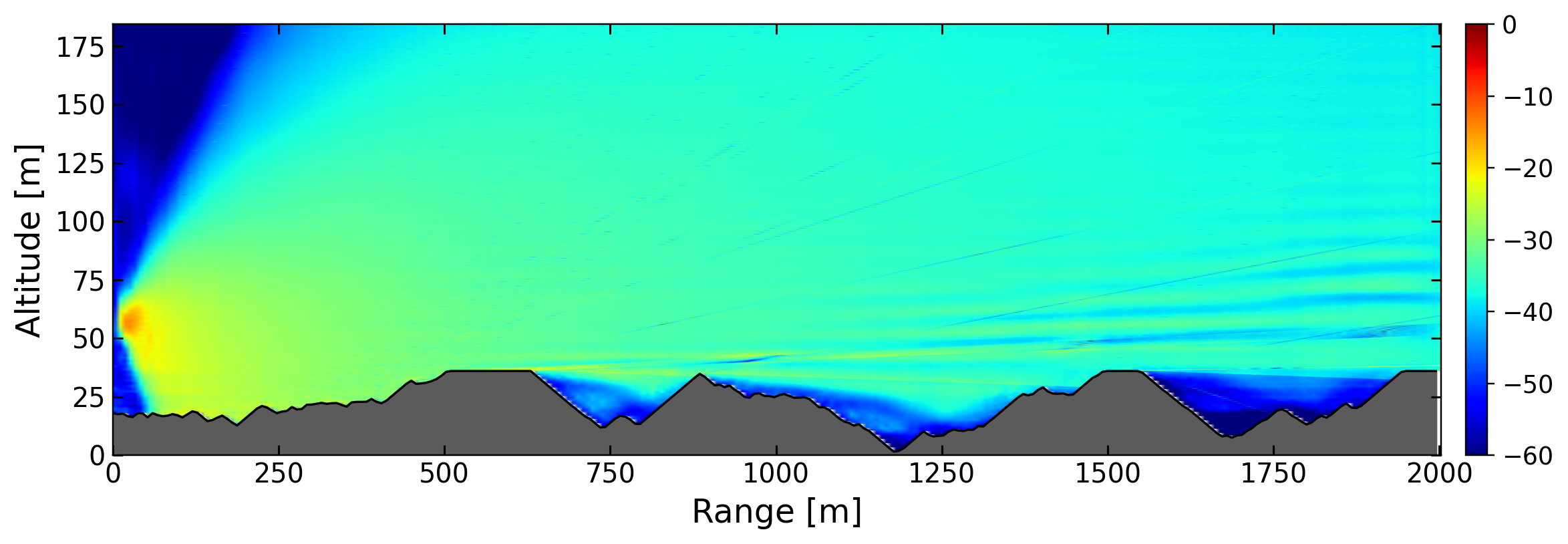}}\\[0.3ex]
        \subfloat[]{\includegraphics[width=0.95\linewidth]{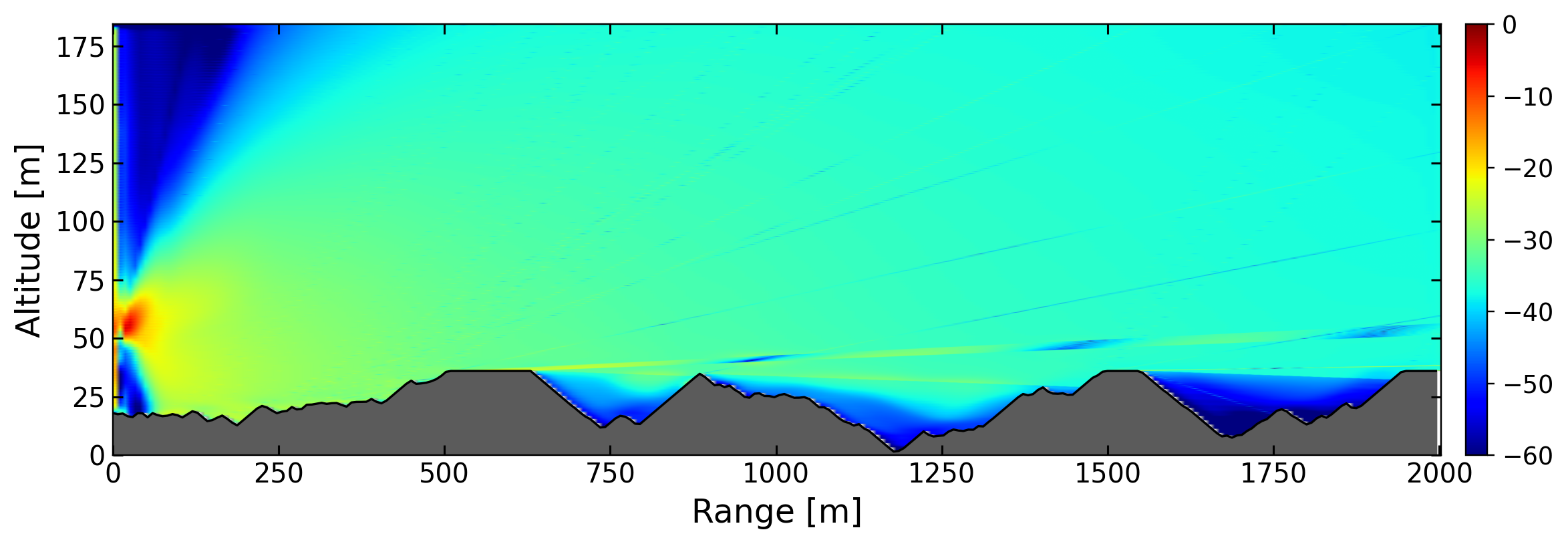}}\\[0.3ex]
        \subfloat[]{\includegraphics[width=0.95\linewidth]{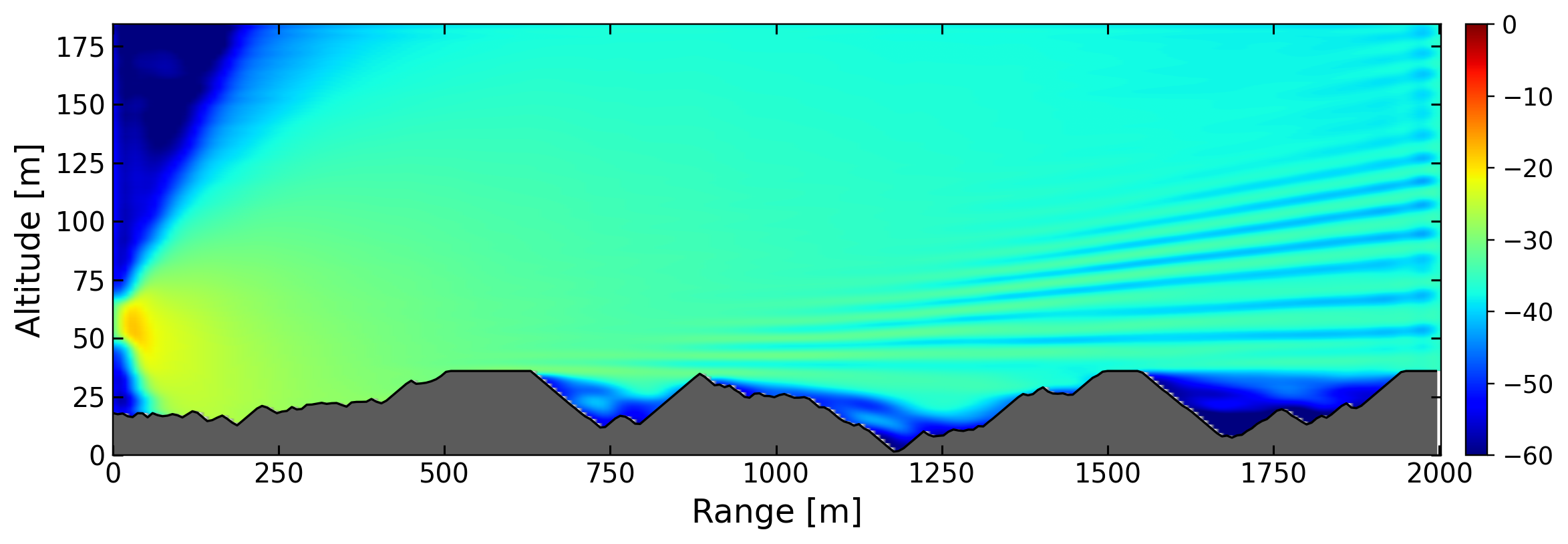}}
    \end{minipage}
    \caption{Representative field-map predictions for the ablation study. (a) SSPE reference. (b) Full model. (c) Without anisotropic convolution. (d) Without antenna conditioning. (e) Without Stage-2 refinement. (f) Without the baseline-derived guidance stack. (g) Without the analytical baseline. The test case uses antenna height $25$~m, 3-dB beamwidth $30^\circ$, and downtilt $10^\circ$.}
    \label{fig:arch_ablation_components}
\end{figure*}

This section evaluates the proposed framework on SSPE-generated
synthetic data. The evaluation includes component ablations, comparison
with a standard U-Net and the reproduced physics-guided CNN of
Huang~\textit{et al.}~\cite{huang2025generalizable}, and a zero-shot
cross-frequency test. The synthetic simulation domain spans $2$~km in
range and $150$~m in height. The main training and test data are
generated at $1.0$~GHz, with $\Delta x=25\lambda=7.5$~m in range and
$\Delta z=0.5\lambda=0.15$~m in height, following the guideline
in~\cite{guo2017pade}. Antenna beamwidth, downtilt, and height are
sampled independently from the discrete grids in
Table~\ref{tab:sspe_params}, and the terrain-trend angle is drawn from
the listed interval.

\begin{table}[!t]
\centering
\caption{Setup of the input parameters for the SSPE simulation.}
\label{tab:sspe_params}
\renewcommand{\arraystretch}{1.2}
\setlength{\tabcolsep}{5pt}
\footnotesize
\begin{tabular}{|c|c|c|c|}
\hline
Parameter & Min. value & Max. value & Increment \\
\hline
Beamwidth [deg] & 15 & 30 & 1 \\
\hline
Downtilt [deg] & 0 & 15 & 1 \\
\hline
Antenna Height [m] & 7 & 35 & 1 \\
\hline
Terrain Trend [deg] & $-0.3$ & $0.3$ & 0.06 \\
\hline
\end{tabular}
\end{table}

Synthetic terrain profiles are generated by superposing $K=50$
random sinusoidal components with logarithmically spaced spatial
frequencies and power-law amplitudes. The frequency of the $k$th
component is defined as
\begin{equation}
f_k = f_{\min} \cdot \left(\frac{f_{\max}}{f_{\min}}\right)^{\!\frac{k-1}{K-1}},
\quad k = 1,\ldots,K,
\label{eq:freq_spacing}
\end{equation}
where $f_{\min}=0.001$~m$^{-1}$ and $f_{\max}=1/(2\Delta x)$ is the
Nyquist limit imposed by the range grid. The corresponding amplitude
follows a $1/f$ power-law spectrum,
\begin{equation}
a_k = a_{\max} \cdot f_k^{-\beta},
\label{eq:amplitude}
\end{equation}
with $a_{\max}=10$ and $\beta=1$, so that large-scale undulations
are emphasized while unrealistically high-frequency oscillations are
suppressed. The raw terrain perturbation is then synthesized as
\begin{equation}
PN(x) = \sum_{k=1}^{K} a_k \sin(2\pi f_k x + \phi_k),
\label{eq:perlin}
\end{equation}
where the phases $\phi_k \sim \mathcal{U}[0,2\pi)$ are drawn
independently for each profile.

To control terrain roughness independently of the spectral shape,
$PN(x)$ is normalized to zero mean and unit variance and then scaled by
a random roughness factor $\eta \sim \mathcal{U}[0.05,0.50]$.
A linear trend with slope angle $\theta$ is subsequently superimposed,
yielding the final elevation profile
\begin{equation}
y(x) = h_0 + 30\eta\,\widetilde{PN}(x) + \tan(\theta)\,x,
\label{eq:terrain}
\end{equation}
where $h_0=18$~m is the transmitter-side ground elevation and
$\widetilde{PN}(x)$ denotes the normalized terrain perturbation. The
resulting profile is clipped to $[0.1,36]$~m, and the elevation change
between adjacent range samples is limited to $1.8$~m in order to avoid
unrealistically steep gradients.

For each valid terrain--antenna configuration, the SSPE solver produces
a two-dimensional field map, which is converted to decibel magnitude
and floored at $-200$~dB. A total of $9{,}836$ samples are retained and
partitioned into training and test sets at a ratio of $8{:}2$. Each
synthetic sample takes an average of $289.77$~s of SSPE simulation time
on a single CPU core. Neural-network inference times are measured on a
workstation equipped with an Intel Xeon Gold~6458Q CPU and an NVIDIA
GeForce RTX~4090~D GPU.

%%%%%%%%%%%%%%%%%%%%%%%%%%%%%%%%%%%%%%%%
\subsection{Ablation on the Proposed Framework}
\label{subsec:arch_ablation}

\begin{figure}[!b]
\centering
\includegraphics[width=\columnwidth]{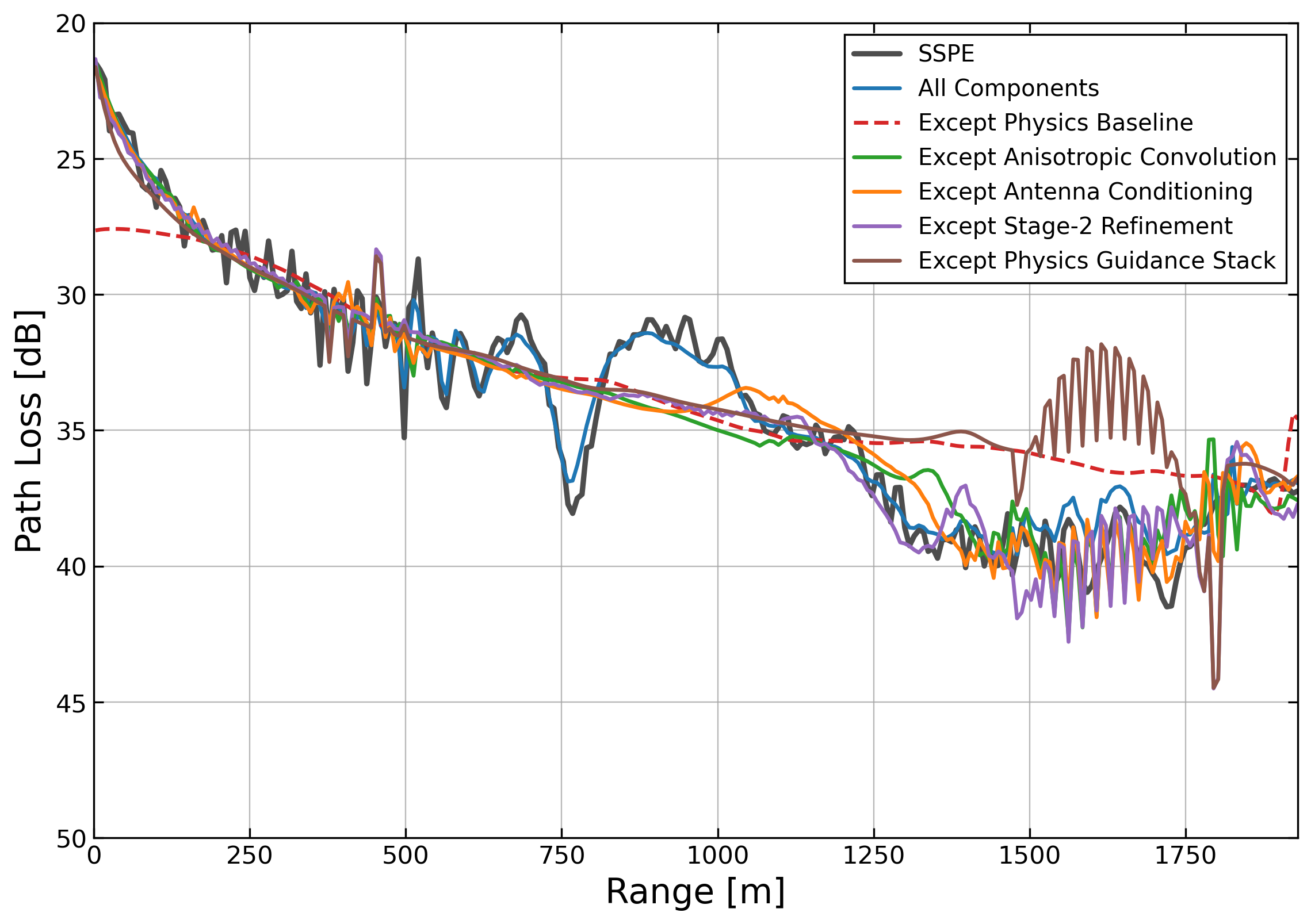}
\caption{Path-loss profiles at $50$~m height for the representative ablation case shown in Fig.~\ref{fig:arch_ablation_components}.}
\label{fig:arch_ablation_pathloss}
\end{figure}

We first isolate the contribution of each design component through an
ablation study. The full model is compared with five variants obtained
by removing anisotropic convolution, antenna conditioning, Stage-2
refinement, the baseline-derived guidance stack, or the analytical
baseline. All variants are trained under the same protocol within this
component-isolation experiment, so the ablation results should be read
as relative comparisons within Table~\ref{tab:arch_ablation}.
Fig.~\ref{fig:arch_ablation_components} shows representative field maps
for a test case with antenna height $25$~m, 3-dB beamwidth $30^\circ$,
and downtilt $10^\circ$; Fig.~\ref{fig:arch_ablation_pathloss} shows
the corresponding path-loss profiles at $50$~m height.

\begin{table}[!t]
\centering
\caption{Ablation study of the proposed framework on the synthetic test set. Lower MAE is better. The symbol ``--'' indicates that the corresponding component is undefined under that ablation.}
\label{tab:arch_ablation}
\setlength{\tabcolsep}{1.6pt}
\renewcommand{\arraystretch}{1.22}
\footnotesize
\begin{tabular}{|c|c|c|c|c|c|}
\hline
\makecell[c]{Anisotropic \\ Convolution} &
\makecell[c]{Antenna \\ Conditioning} &
\makecell[c]{Stage-2 \\ Refinement} &
\makecell[c]{Baseline-Derived \\ Guidance \\ Stack} &
\makecell[c]{Analytical \\ Baseline} &
\makecell[c]{MAE \\ (dB)} \\
\hline
\checkmark & \checkmark & \checkmark & \checkmark & \checkmark & 2.2835 \\
\hline
$\times$   & \checkmark & \checkmark & \checkmark & \checkmark & 4.1828 \\
\hline
\checkmark & $\times$   & \checkmark & \checkmark & \checkmark & 3.7282 \\
\hline
\checkmark & \checkmark & $\times$   & --         & \checkmark & 3.4892 \\
\hline
\checkmark & \checkmark & \checkmark & $\times$   & \checkmark & 6.9426 \\
\hline
\checkmark & \checkmark & \checkmark & --         & $\times$   & 2.9837 \\
\hline
\end{tabular}
\end{table}
In this ablation run, the full configuration gives the lowest MAE,
$2.2835$~dB. Removing the baseline-derived guidance stack causes the largest
degradation, increasing the MAE to $6.9426$~dB. The corresponding field
map in Fig.~\ref{fig:arch_ablation_components}(f) shows weaker recovery
of the terrain-induced shadow region and post-obstacle field recovery,
which indicates that the baseline-derived guidance channels provide an
important prior for residual correction. Removing anisotropic
convolution increases the MAE to $4.1828$~dB. In
Fig.~\ref{fig:arch_ablation_components}(c), the prediction is smoother
along the propagation direction and loses part of the oblique fringe
continuity, supporting the use of range-aligned kernels for the SSPE
grid. Removing antenna conditioning increases the MAE to $3.7282$~dB;
the prediction in Fig.~\ref{fig:arch_ablation_components}(d) still
captures the coarse field structure but gives a less accurate
large-scale attenuation trend. Without Stage~2 refinement, the MAE
rises to $3.4892$~dB, and
Fig.~\ref{fig:arch_ablation_components}(e) shows reduced fine-scale
oscillatory detail and weaker local shadow transitions. The variant
without the analytical baseline is different from the other ablations
because the network then performs full-field regression and the
baseline-derived guidance stack is unavailable. Its MAE increases to $2.9837$~dB, with
weaker fringe recovery and less accurate shadow transitions than the
full model.

The one-dimensional comparison in
Fig.~\ref{fig:arch_ablation_pathloss} follows the same trend. The full
model tracks the SSPE path-loss curve most closely, whereas each
ablated variant introduces a distinct deviation pattern. Removing the
baseline-derived guidance stack or anisotropic convolution gives the largest
errors, while the other three ablations produce smaller but still
visible degradations. These results support the contribution of the
baseline-derived guidance, anisotropic feature extraction, antenna
conditioning, and two-stage refinement.

\subsection{Comparison with Learning-Based Methods}
\label{subsec:arch_comparison}

\begin{figure}[!t]
    \centering
    \subfloat[]{\includegraphics[width=\columnwidth]{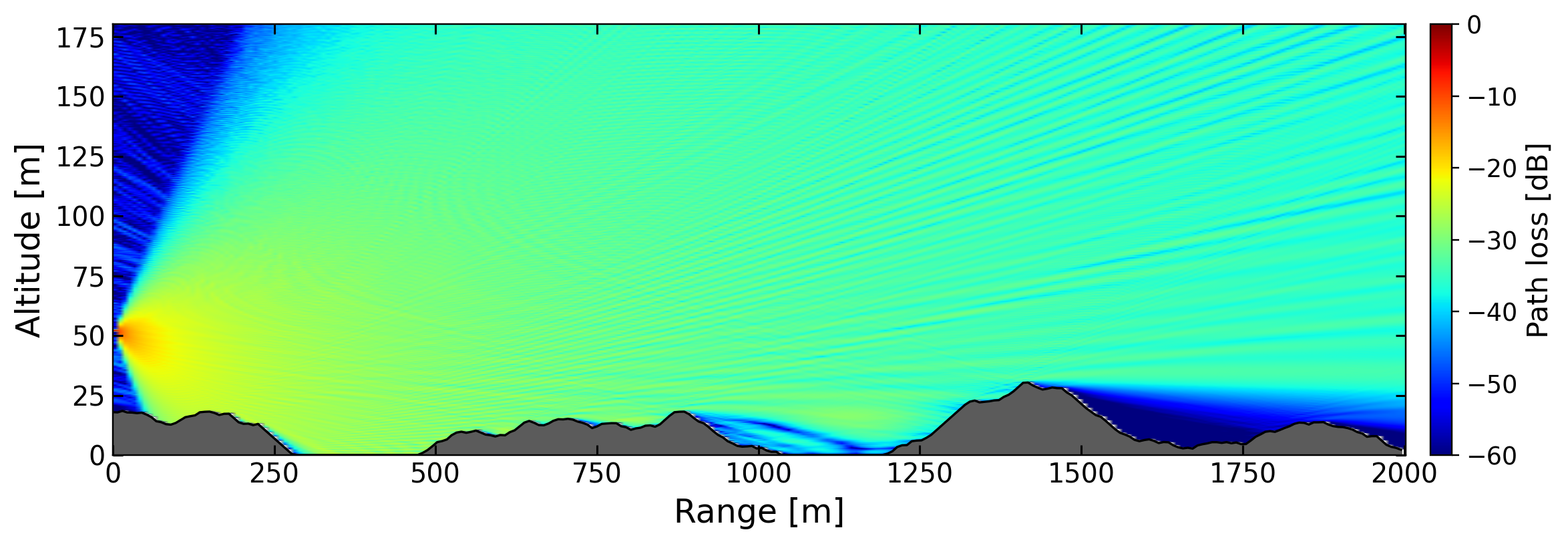}}\\[0.3ex]
    \subfloat[]{\includegraphics[width=\columnwidth]{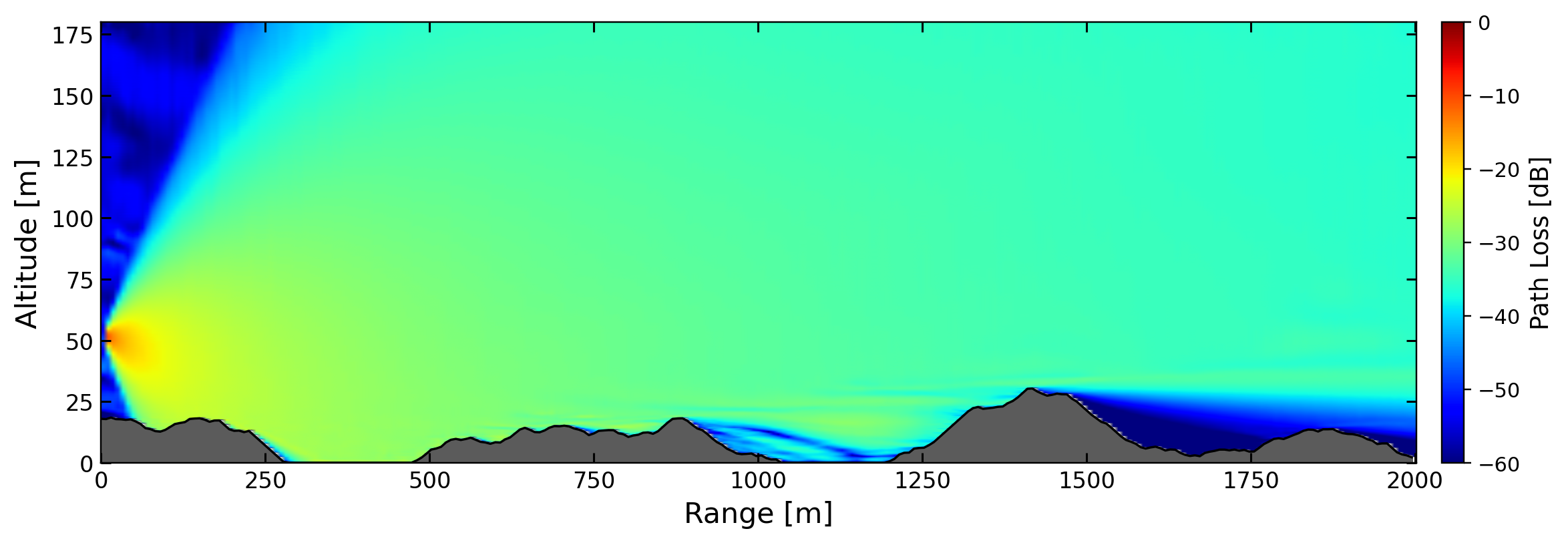}}\\[0.3ex]
    \subfloat[]{\includegraphics[width=\columnwidth]{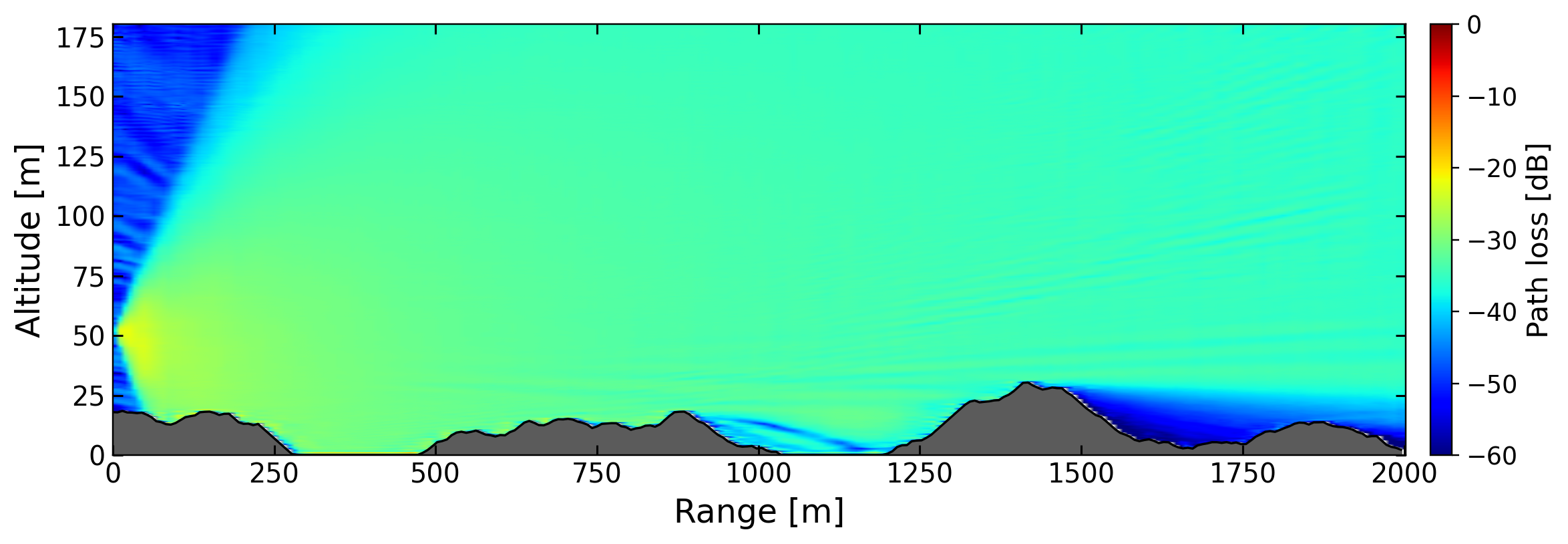}}\\[0.3ex]
    \subfloat[]{\includegraphics[width=\columnwidth]{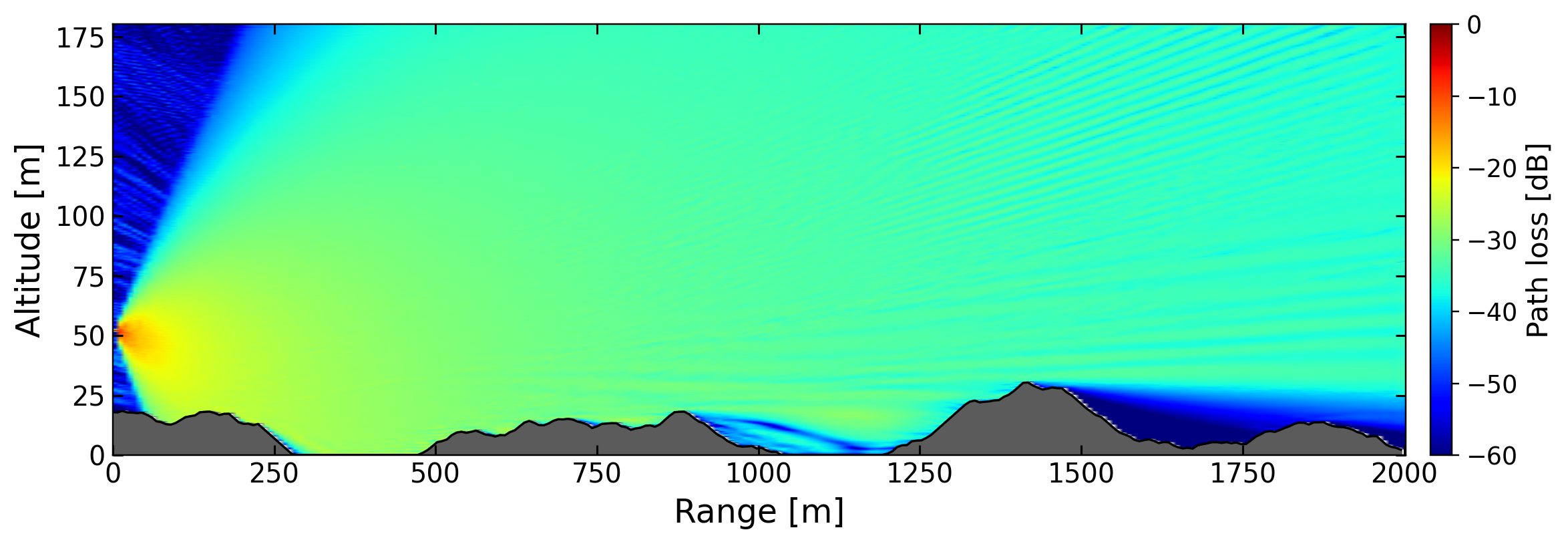}}
    \caption{Representative field-map comparison among learning-based
methods. (a) SSPE reference. (b) Standard U-Net. (c) Huang's
model~\cite{huang2025generalizable}. (d) Proposed model. The test
case uses antenna height $16$~m, 3-dB beamwidth $28^\circ$, and
downtilt $6^\circ$.}
    \label{fig:comparison_maps}
\end{figure}

\begin{figure}[!b]
\centering
\includegraphics[width=\columnwidth]{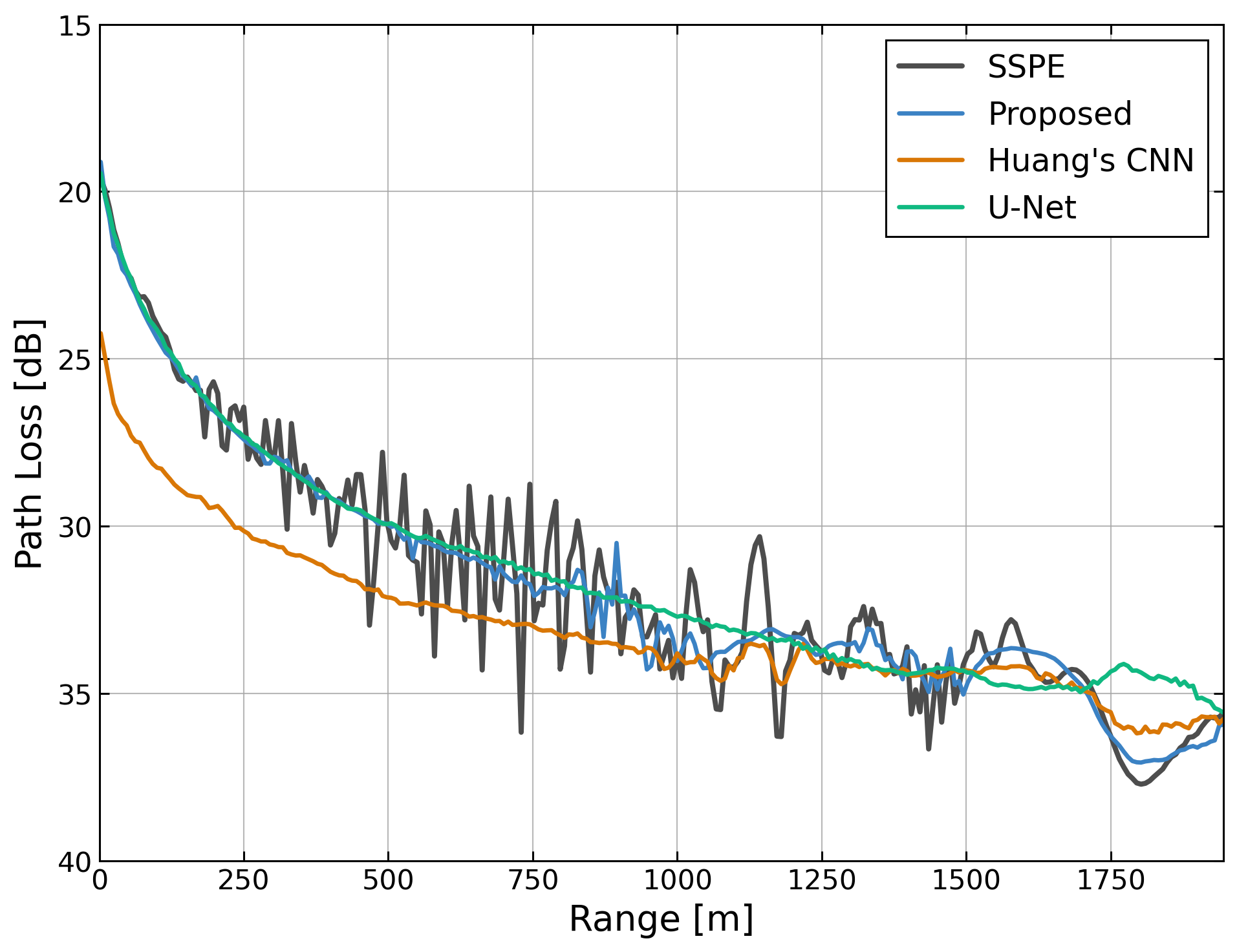}
\caption{Path-loss profiles at $50$~m height for the representative
comparison case shown in Fig.~\ref{fig:comparison_maps}.}
\label{fig:comparison_pathloss}
\end{figure}

We next compare the proposed framework with two learning-based
surrogates on the same held-out synthetic test set. The first is a
standard U-Net trained to regress the full SSPE field from terrain and
antenna inputs without a physics prior. The second is the reproduced
physics-guided CNN of Huang~\textit{et al.}~\cite{
huang2025generalizable}. These two comparators separate the value of
adding physics-related inputs from the value of the proposed
analytical-baseline residual formulation. All three methods are
evaluated under the same test protocol. Table~\ref{tab:main_comparison}
reports the quantitative results, and Figs.~\ref{fig:comparison_maps}
and~\ref{fig:comparison_pathloss} show a representative field map and
path-loss profile.

\begin{table}[!t]
\centering
\caption{Comparison with learning-based methods on the same held-out
synthetic test set. Lower MAE and lower inference time are better.}
\label{tab:main_comparison}
\renewcommand{\arraystretch}{1.2}
\setlength{\tabcolsep}{3pt}
\footnotesize
\begin{tabular}{|c|c|c|c|}
\hline
Method & MAE (dB) & \makecell[c]{Avg.\ Inference \\ (ms/sample)} & \# Parameters \\
\hline
Standard U-Net & 3.3955 & 45.75 & 17,325,697 \\
\hline
Huang et al.~\cite{huang2025generalizable} & 3.2560 & 27.30 & 44,603,399 \\
\hline
Proposed & 2.1061 & 11.30 & 8,207,266 \\
\hline
\end{tabular}
\end{table}

Table~\ref{tab:main_comparison} shows that the proposed model improves
both accuracy and efficiency on this test set. The standard U-Net gives
an MAE of $3.3955$~dB, and the reproduced model of
Huang~\textit{et al.} gives an MAE of $3.2560$~dB. The proposed model
reduces the MAE to $2.1061$~dB, corresponding to reductions of
$37.9\%$ and $35.3\%$ relative to these two baselines, respectively.
It also uses $8.2$M parameters and an average inference time of
$11.3$~ms per sample, compared with $17.3$M parameters and $45.75$~ms
for the standard U-Net and $44.6$M parameters and $27.30$~ms for
Huang's model. These results are consistent with the intended role of
the hybrid analytical-neural formulation: the deterministic baseline
absorbs part of the large-scale propagation trend, leaving the neural
network to model the terrain-dependent correction.

The qualitative comparison in Fig.~\ref{fig:comparison_maps} is
consistent with the quantitative results. The standard U-Net captures
the gross attenuation pattern but smooths much of the terrain-coupled
fringe structure and misses some local shadow transitions. Huang's
model improves the large-scale envelope but remains relatively smooth
in the fine interference pattern. The proposed model better preserves
both the attenuation envelope and the oscillatory fringe structure in
this representative case.

The one-dimensional path-loss comparison in
Fig.~\ref{fig:comparison_pathloss} gives the same picture. The proposed
prediction follows the SSPE reference more closely over the $2$-km
range, while the standard U-Net and Huang's model show larger
deviations in the attenuation trend and local oscillatory detail. These
comparisons suggest that the analytical-baseline residual formulation
is more effective for this task than adding physics conditioning to a
direct-regression model alone.

% \FloatBarrier % Removed to allow floats to be placed more flexibly
%%%%%%%%%%%%%%%%%%%%%%%%%%%%%%%%%%%%%%%%
\subsection{Cross-Frequency Generalization}
\label{subsec:generalization}

An important practical requirement for propagation surrogates is the
ability to remain useful when the carrier frequency changes. In
conventional fixed-frequency surrogates, frequency dependence is learned
implicitly from the training data, so extending the model to another
band typically requires new training data and retraining. In the
proposed framework, the dominant frequency dependence is represented
explicitly in the deterministic baseline $B$, while the neural network
learns a residual correction associated with the terrain and antenna
configuration. This structure allows the trained residual model to be
tested at an unseen frequency by recomputing only the analytical
baseline.

To assess this behavior, the model is trained only on the synthetic
$1.0$~GHz dataset and then evaluated directly on a held-out
$1.8$~GHz test set containing $150$ samples. No $1.8$~GHz samples are
used during training or fine-tuning. At test time, the analytical
baseline is recomputed at $1.8$~GHz, while all learned parameters are
kept fixed.
This experiment is intended to test whether the residual model remains
useful for one unseen nearby frequency after baseline recomputation; it
is not intended as evidence of broadband generalization.

\begin{figure}[!t]
    \centering
    \subfloat[]{\includegraphics[width=\columnwidth]{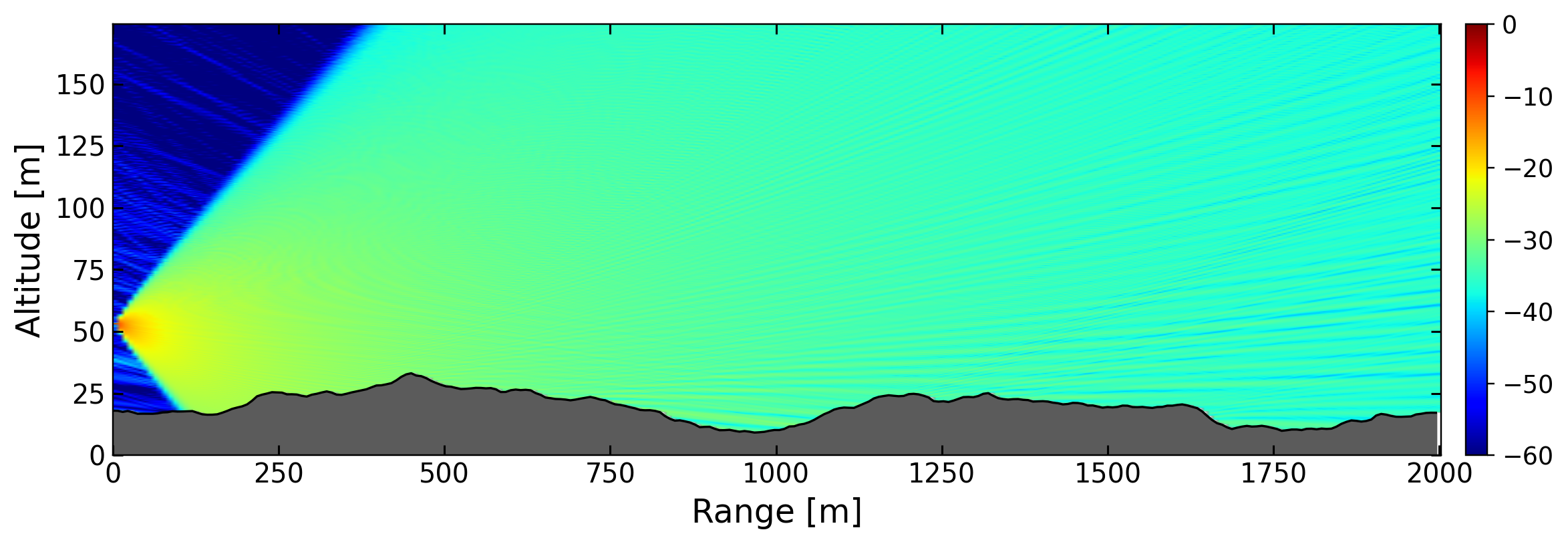}}\\[0.3ex]
    \subfloat[]{\includegraphics[width=\columnwidth]{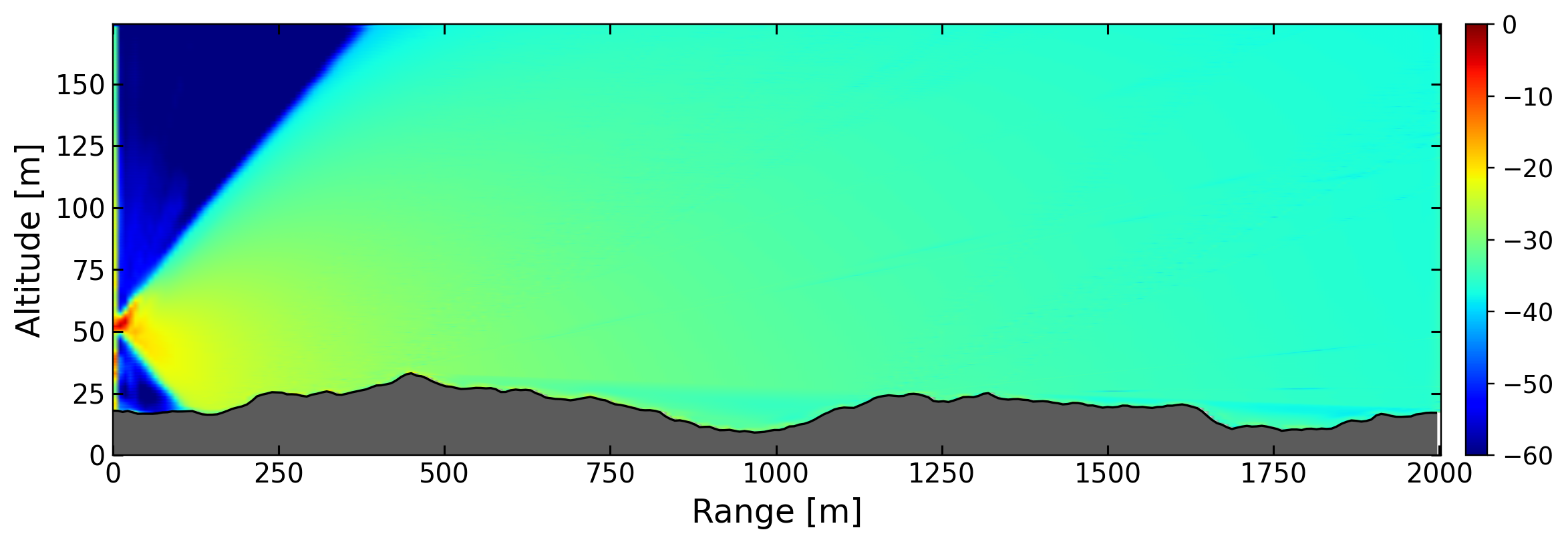}}
    \caption{Representative field-map prediction for zero-shot transfer to $1.8$~GHz. (a) SSPE reference. (b) Proposed zero-shot prediction. The test case uses antenna height $26$~m, 3-dB beamwidth $25^\circ$, and downtilt $4^\circ$.}
    \label{fig:freq_transfer_maps}
\end{figure}

\begin{figure}[!b]
\centering
\includegraphics[width=\columnwidth]{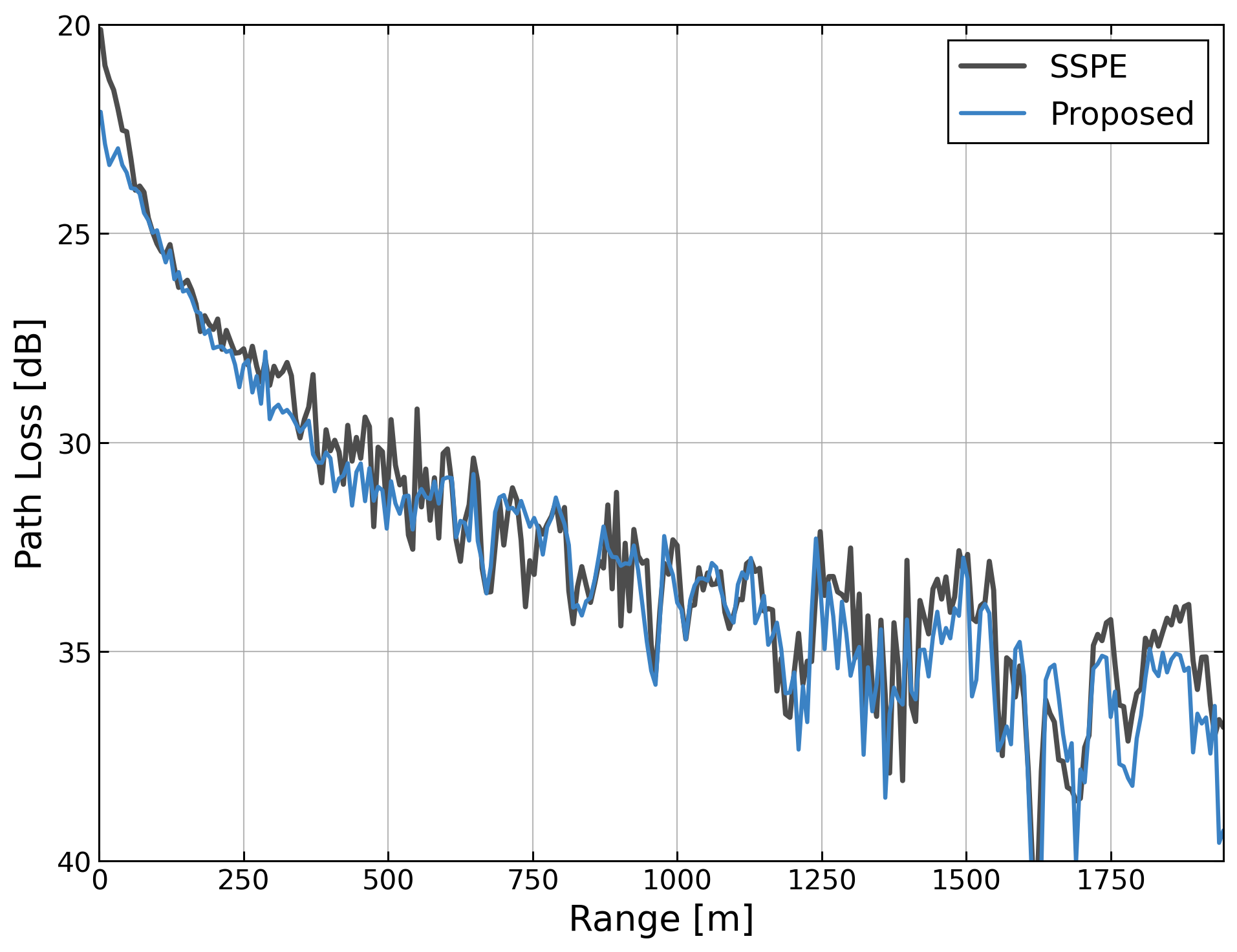}
\caption{Path-loss profiles at $50$~m height for the representative zero-shot transfer case shown in Fig.~\ref{fig:freq_transfer_maps}.}
\label{fig:freq_transfer_pathloss}
\end{figure}

The qualitative field-map comparison in
Fig.~\ref{fig:freq_transfer_maps} shows that the zero-shot prediction
retains the dominant attenuation pattern and the main terrain-coupled
shadow transition at the unseen carrier frequency. The path-loss curve
in Fig.~\ref{fig:freq_transfer_pathloss} follows the SSPE reference over
most of the range, although local deviations remain visible. The MAE on
the $1.8$~GHz test set is $2.8692$~dB, compared with $2.1061$~dB for
the in-frequency test at $1.0$~GHz. This increase is expected under a
substantial frequency shift, but the result indicates that the residual
model still carries useful information when the analytical baseline is
recomputed at the new frequency.

%%%%%%%%%%%%%%%%%%%%%%%%%%%%%%%%%%%%%%%%
\section{Validation on a Measured Terrain Profile}
\label{sec:realistic_irregular_terrains}

\begin{figure}[!t]
    \centering
    % Show the measured and predicted Hjorringvej field maps side by side to
    % limit the vertical extent of this figure.
    \subfloat[]{\includegraphics[width=\columnwidth]{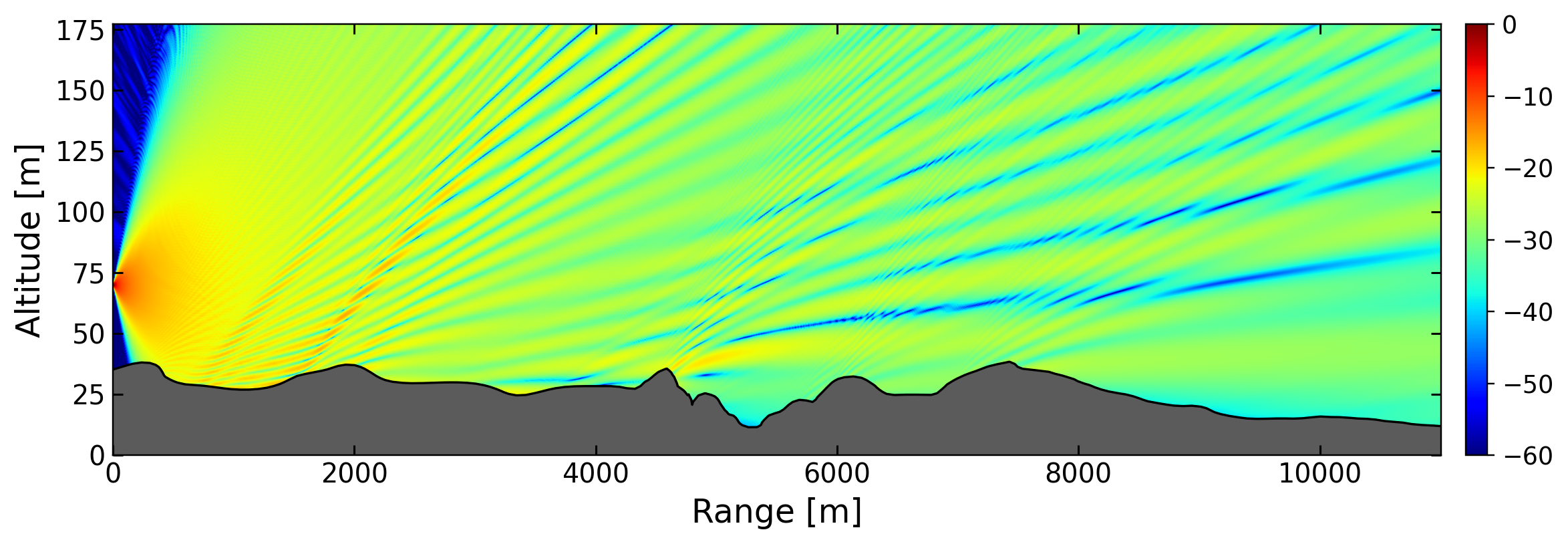}\label{fig:hjorringvej_gt}}\\[0.3ex]
    \subfloat[]{\includegraphics[width=\columnwidth]{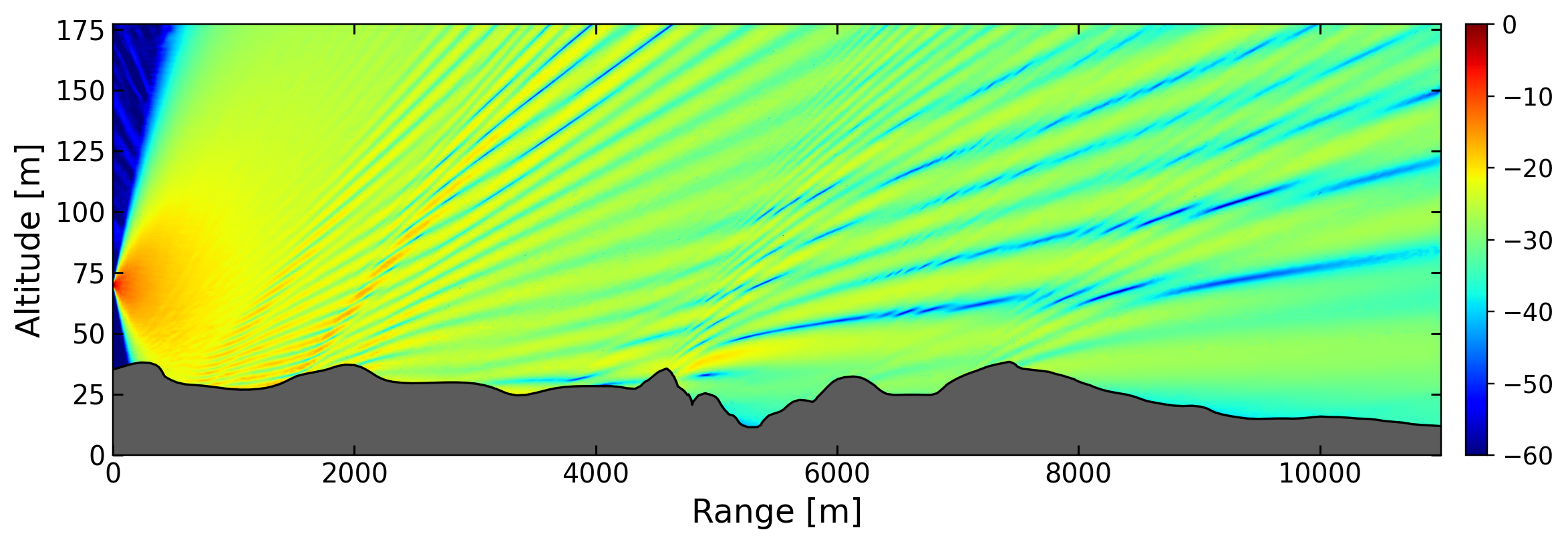}\label{fig:hjorringvej_pred}}
    \caption{Field-map comparison for the measured \textit{Hjorringvej} terrain profile at $970$~MHz. (a) SSPE reference. (b) Proposed model prediction.}
    \label{fig:hjorringvej_maps}
\end{figure}

\begin{figure}[!b]
\centering
\includegraphics[width=\columnwidth]{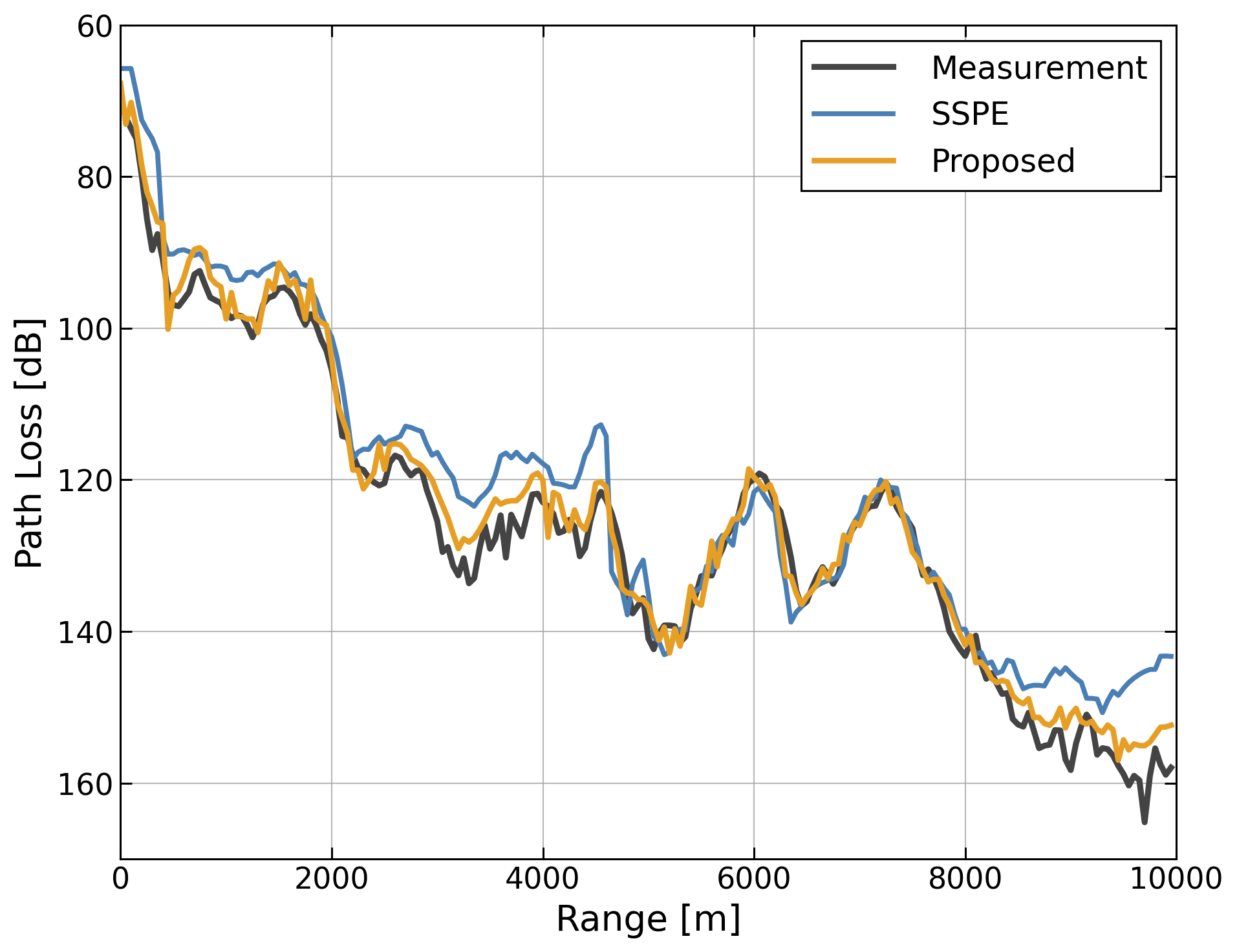}
\caption{Path-loss comparison for the measured \textit{Hjorringvej} terrain profile at $970$~MHz. The curves correspond to the field measurements, the SSPE reference, and the proposed model.}
\label{fig:hjorringvej_pathloss}
\end{figure}

The proposed model is also tested on the measured Hjorringvej irregular
terrain profile in northern Denmark~\cite{hviid1995terrain}. This case
provides an external profile that is not drawn from the synthetic
terrain generator used for training. The profile spans $11$~km and is
evaluated at $970$~MHz over a lossy dielectric ground with relative
permittivity $\varepsilon_r=4.5$ and loss tangent $\tan\delta=0.07$.
Figs.~\ref{fig:hjorringvej_maps} and~\ref{fig:hjorringvej_pathloss}
compare the proposed prediction with the SSPE reference and the
available field measurements.

The field-map comparison in Fig.~\ref{fig:hjorringvej_maps} shows that
the proposed prediction captures the dominant attenuation trend and the
main terrain-induced shadowing pattern along the route. The path-loss
comparison in Fig.~\ref{fig:hjorringvej_pathloss} further shows that
the proposed curve remains close to the SSPE reference over much of the
profile and broadly follows the measured trend. This result should be
viewed as a measured-profile case study rather than a complete
measurement campaign. It provides an external consistency check on a
real terrain profile, but broader measurement-based validation remains
necessary. For this profile, SSPE takes $9{,}343$~s to
simulate the full field, whereas the proposed model takes $0.018$~s.

% =========================
% Section VI: Conclusion
% =========================
\section{Conclusion}
This paper presented a hybrid analytical-neural residual surrogate for
fast radio propagation prediction over irregular terrain. The method
separates the SSPE field into a deterministic analytical baseline and a
learned residual correction. The baseline represents free-space
spreading, antenna directivity, modified knife-edge diffraction, and
terrain-aware specular reflection, while the neural network models the
remaining terrain-dependent mismatch through a lightweight
coarse-to-fine architecture.

Experiments on SSPE-generated synthetic data showed that this
analytical-baseline residual formulation improves the
accuracy-efficiency tradeoff relative to direct-regression learning
baselines. On the main synthetic test set, the proposed model reduced
the MAE from $3.3955$~dB for a standard U-Net and $3.2560$~dB for a
reproduced physics-guided CNN to $2.1061$~dB, while using fewer
parameters and lower inference time. The component ablations further
showed that the baseline-derived guidance stack, anisotropic
convolutions, antenna conditioning, Stage-2 refinement, and analytical
baseline each contribute to the final prediction quality.

The present results also suggest the practical value of keeping
frequency dependence in the analytical baseline. A model trained at
$1.0$~GHz achieved an MAE of $2.8692$~dB when evaluated at $1.8$~GHz
without retraining, after recomputing only the baseline. In addition,
the Hjorringvej case study showed that the proposed surrogate remains
close to the SSPE reference and broadly follows the measured path-loss
trend on an external terrain profile. The current validation is still
limited to one unseen frequency and one measured profile. Future work
will extend the evaluation to broader frequency ranges, more diverse
terrain classes, and larger measurement-based test sets.

\bibliographystyle{ieeetr}
%\bibliography{IEEEabrv./MyBibliography}
\bibliography{MyBibliography}

\begin{thebibliography}{10}

\bibitem{guo2017pade}
Q.~Guo and Y.~Long, ``Pade second-order parabolic equation modeling for
  propagation over irregular terrain,'' {\em IEEE Antennas and Wireless
  Propagation Letters}, vol.~16, pp.~2852--2855, 2017.

\bibitem{ozgun2009recursive}
O.~Ozgun, ``Recursive two-way parabolic equation approach for modeling terrain
  effects in tropospheric propagation,'' {\em IEEE Transactions on Antennas and
  Propagation}, vol.~57, no.~9, pp.~2706--2714, 2009.

\bibitem{zhang2016applying}
P.~Zhang, L.~Bai, Z.~Wu, and L.~Guo, ``Applying the parabolic equation to
  tropospheric groundwave propagation: A review of recent achievements and
  significant milestones,'' {\em IEEE Antennas and Propagation Magazine},
  vol.~58, no.~3, pp.~31--44, 2016.

\bibitem{qin2024comparative}
H.~Qin and X.~Zhang, ``Comparative analysis of finite-difference and split-step
  based parabolic equation methods for tunnel propagation modelling,'' {\em IET
  Microwaves, Antennas \& Propagation}, vol.~18, no.~2, pp.~59--72, 2024.

\bibitem{seretis2022toward}
A.~Seretis and C.~D. Sarris, ``Toward physics-based generalizable convolutional
  neural network models for indoor propagation,'' {\em IEEE Transactions on
  Antennas and Propagation}, vol.~70, no.~6, pp.~4112--4126, 2022.

\bibitem{seretis2023generalizable}
A.~Seretis, V.~Jevremovic, A.~Jemmali, and C.~D. Sarris, ``Generalizable
  machine-learning-based modeling of radiowave propagation in stadiums,'' {\em
  IEEE Open Journal of Antennas and Propagation}, vol.~4, pp.~1116--1128, 2023.

\bibitem{bakirtzis2022deepray}
S.~Bakirtzis, J.~Chen, K.~Qiu, J.~Zhang, and I.~Wassell, ``Em deepray: An
  expedient, generalizable, and realistic data-driven indoor propagation
  model,'' {\em IEEE Transactions on Antennas and Propagation}, vol.~70, no.~6,
  pp.~4140--4154, 2022.

\bibitem{9210205}
Z.~Ma, K.~Xu, R.~Song, C.-F. Wang, and X.~Chen, ``Learning-based fast
  electromagnetic scattering solver through generative adversarial network,''
  {\em IEEE Transactions on Antennas and Propagation}, vol.~69, no.~4,
  pp.~2194--2208, 2021.

\bibitem{9311729}
V.~V. Ratnam, H.~Chen, S.~Pawar, B.~Zhang, C.~J. Zhang, Y.-J. Kim, S.~Lee,
  M.~Cho, and S.-R. Yoon, ``Fadenet: Deep learning-based mm-wave large-scale
  channel fading prediction and its applications,'' {\em IEEE Access}, vol.~9,
  pp.~3278--3290, 2021.

\bibitem{10057211}
T.~Shan, J.~Zeng, X.~Song, R.~Guo, M.~Li, F.~Yang, and S.~Xu,
  ``Physics-informed supervised residual learning for electromagnetic
  modeling,'' {\em IEEE Transactions on Antennas and Propagation}, vol.~71,
  no.~4, pp.~3393--3407, 2023.

\bibitem{10065920}
H.~Qin and X.~Zhang, ``Two-way parabolic equation method for millimeter-wave
  propagation modeling in tunnels,'' in {\em 2022 IEEE 8th International
  Conference on Computer and Communications (ICCC)}, pp.~1573--1577, 2022.

\bibitem{huang2024tunnel}
S.~Huang, S.~Wang, and X.~Zhang, ``Generalisable convolutional neural network
  model for radio wave propagation in tunnels,'' {\em IET Microwaves, Antennas
  \& Propagation}, vol.~18, no.~7, pp.~467--479, 2024.

\bibitem{brennan2023site}
C.~Brennan and K.~McGuinness, ``Site-specific deep learning path loss models
  based on the method of moments,'' in {\em 2023 17th European Conference on
  Antennas and Propagation (EuCAP)}, pp.~1--5, IEEE, 2023.

\bibitem{huang2025generalizable}
S.~Huang, H.~Qin, W.~Hou, X.~Zhang, and X.~Zhang, ``A generalizable
  physics-guided convolutional neural network for irregular terrain
  propagation,'' {\em IEEE Transactions on Antennas and Propagation}, vol.~73,
  no.~6, pp.~3975--3985, 2025.

\bibitem{zhang2020cellular}
X.~Zhang, J.~Ren, X.~Shu, L.~Zhou, B.~Zhang, and X.~Chen, ``Cellular network
  radio propagation modeling with deep convolutional neural networks,'' in {\em
  Proceedings of the 26th ACM SIGKDD International Conference on Knowledge
  Discovery \& Data Mining}, pp.~2378--2386, 2020.

\bibitem{liu2021deeplora}
L.~Liu, Y.~Yao, Z.~Cao, and M.~Zhang, ``{DeepLoRa}: Learning accurate path loss
  model for long distance links in {LPWAN},'' in {\em IEEE INFOCOM 2021 - IEEE
  Conference on Computer Communications}, pp.~1--10, 2021.

\bibitem{levie2021radiounet}
R.~Levie, {\c{C}}.~Yapar, G.~Kutyniok, and G.~Caire, ``{RadioUNet}: Fast radio
  map estimation with convolutional neural networks,'' {\em IEEE Transactions
  on Wireless Communications}, vol.~20, no.~6, pp.~4001--4015, 2021.

\bibitem{chen2021distributed}
M.~Chen, D.~G{\"u}nd{\"u}z, K.~Huang, W.~Saad, M.~Bennis, A.~V. Feljan, and
  H.~V. Poor, ``Distributed learning in wireless networks: Recent progress and
  future challenges,'' {\em IEEE J. Sel. Areas Commun.}, vol.~39, no.~12,
  pp.~3579--3605, 2021.

\bibitem{kim2016accurate}
J.~Kim, J.~K. Lee, and K.~M. Lee, ``Accurate image super-resolution using very
  deep convolutional networks,'' in {\em Proceedings of the IEEE Conference on
  Computer Vision and Pattern Recognition (CVPR)}, pp.~1646--1654, 2016.

\bibitem{zhang2017beyond}
K.~Zhang, W.~Zuo, Y.~Chen, D.~Meng, and L.~Zhang, ``Beyond a {Gaussian}
  denoiser: Residual learning of deep {CNN} for image denoising,'' {\em IEEE
  Transactions on Image Processing}, vol.~26, no.~7, pp.~3142--3155, 2017.

\bibitem{alrabeiah2019deep}
M.~Alrabeiah and A.~Alkhateeb, ``Deep learning for {TDD} and {FDD} massive
  {MIMO}: Mapping channels in space and frequency,'' in {\em 2019 53rd Asilomar
  Conference on Signals, Systems, and Computers}, pp.~1465--1470, 2019.

\bibitem{sung2025commUNext}
C.-J. Sung, F.-H. Lin, T.-H. Huang, C.-H. Huang, H.~Chen, C.-K. Wen, and
  H.~Wymeersch, ``{CommUNext}: Deep learning-based cross-band and
  multi-directional signal prediction,'' 2025.

\bibitem{levy2000parabolic}
M.~Levy, {\em Parabolic Equation Methods for Electromagnetic Wave Propagation}.
\newblock London, U.K.: Institution of Electrical Engineers, 2000.

\bibitem{apaydin2017radio}
G.~Apaydin and L.~Sevgi, {\em Radio Wave Propagation and Parabolic Equation
  Modeling}.
\newblock Hoboken, NJ, USA: Wiley-IEEE Press, 2017.

\bibitem{boscaini2016learning}
D.~Boscaini, J.~Masci, E.~Rodol\`a, and M.~Bronstein, ``Learning shape
  correspondence with anisotropic convolutional neural networks,'' in {\em
  Advances in Neural Information Processing Systems (NeurIPS)}, pp.~3189--3197,
  2016.

\bibitem{yu2017dilated}
F.~Yu, V.~Koltun, and T.~Funkhouser, ``Dilated residual networks,'' in {\em
  Proceedings of the IEEE Conference on Computer Vision and Pattern Recognition
  (CVPR)}, pp.~636--644, 2017.

\bibitem{perez2018film}
E.~Perez, F.~Strub, H.~de~Vries, V.~Dumoulin, and A.~C. Courville, ``{FiLM}:
  Visual reasoning with a general conditioning layer,'' in {\em Proceedings of
  the Thirty-Second AAAI Conference on Artificial Intelligence (AAAI)},
  pp.~3942--3951, 2018.

\bibitem{hviid1995terrain}
J.~T. Hviid, J.~B. Andersen, J.~Toftg{\aa}rd, and J.~Bojer, ``Terrain-based
  propagation model for rural area---an integral equation approach,'' {\em IEEE
  Transactions on Antennas and Propagation}, vol.~43, no.~1, pp.~41--46, 1995.

\end{thebibliography}

%% Or manually put reference below
%\begin{thebibliography}{10}

%\bibitem{Levy00}
%M. Levy, \emph{Parabolic Equation Methods for Electromagnetic Wave Propagation.} London, U.K.: Inst. Elect. Eng., 2000.

%\end{thebibliography}

%%%%%%%%%%%%%%%%%%%%%%%%%%%%%%%%%%%%%%%%%%%%%%%%%%%%%%%%%%%%%%%%%%%%%%%%%%%%%%%%
%%%%%%%%%%%%%%%%%%%%%%%%%%%%%%%%%%%%%%%%%%%%%%%%%%%%%%%%%%%%%%%%%%%%%%%%%%%%%%%%
%% Biography

%%%%%%%%%%%%%%%%%%%%%%%%%%%%%%%%%%%%%%%%%%%%%%%%%%%%%%%%%%%%%%%%%%%%%%%%%%%%%%%%
%%%%%%%%%%%%%%%%%%%%%%%%%%%%%%%%%%%%%%%%%%%%%%%%%%%%%%%%%%%%%%%%%%%%%%%%%%%%%%%%
\end{document}